\documentclass[12pt]{iopart}
\usepackage{epsfig}
\pdfoutput=1
\usepackage{color}
\usepackage{amssymb}

\usepackage{graphicx} 

\def\beqra{\begin{eqnarray}}
\def\eeqra{\end{eqnarray}}
\def\beq{\begin{equation}}
\def\eeq{\end{equation}}

\def\etain{\eta_{in}}
\def\tW{\tilde W}

\def\vp{\varphi}
\def\vpb{\bar\varphi}

\def\bx{{\bf{x}}}
\def\bk{{\bf{k}}}
\def\bp{{\bf{p}}}
\def\bq{{\bf{q}}}

\def\bv{{\bf{v}}}

\def\hmpc{{\mathrm{h\; Mpc}^{-1}}}
\def\mpch{\mathrm{h}^{-1}\,\mathrm{Mpc}}
\def\bV0{{\bf{V_0}}}
\def\re#1{(\ref{#1})}

\def\la{~\mbox{\raisebox{-.6ex}{$\stackrel{<}{\sim}$}}~}
\def\ga{~\mbox{\raisebox{-.6ex}{$\stackrel{>}{\sim}$}}~}

\def\bx{{\bf{x}}}
\def\by{{\bf{y}}}

\def\bk{{\bf{k}}}
\def\bp{{\bf{p}}}
\def\bq{{\bf{q}}}

\def\bv{{\bf{v}}}

\begin{document}
\begin{flushright}
{\small UMN-TH-3339/14}
\end{flushright}
%
\title[A  coarse grained perturbation theory for the Large Scale Structure]{A  coarse grained perturbation theory for the Large Scale Structure,  
with cosmology and time independence in the UV}
\author{Alessandro Manzotti$^{1}$, Marco Peloso$^{2}$, Massimo Pietroni$^{3,4}$, Matteo Viel$^{5,6}$, Francisco Villaescusa-Navarro$^{5,6}$}
\vskip 0.3 cm
\address{
$^1$Kavli Institute for Cosmological Physics, Department of Astronomy \& Astrophysics,
Enrico Fermi Institute, University of Chicago, Chicago, Illinois 60637, U.S.A\\
$^2$School of Physics and Astronomy, University of Minnesota, Minneapolis, 55455, USA\\
$^3$INFN, Sezione di Padova, via Marzolo 8, I-35131, Padova, Italy\\
$^4$ Dipartimento di Fisica e Scienze della Terra, Universit\`a di Parma, Viale Usberti
7/A, I-43100 Parma, Italy\\
$^5$INAF - Osservatorio Astronomico di Trieste, Via G.B. Tiepolo 11, I-34143 Trieste, Italy\\
$^6$INFN, Sezione di Trieste, Via Valerio 2, I-34127 Trieste, Italy
}

\begin{abstract}
Standard cosmological perturbation theory  (SPT)  for the Large Scale Structure (LSS) of the Universe fails at small scales (UV) due to strong nonlinearities and to multistreaming effects. In ref.~\cite{Pietroni:2011iz} a new framework was proposed in which the large scales (IR) are treated perturbatively while the information on the UV, mainly small scale velocity dispersion, is obtained by nonlinear methods like N-body simulations. 
Here we develop this approach, showing that it is possible to reproduce the fully nonlinear power spectrum (PS) by combining a simple (and fast) 1-loop computation for the IR scales and the measurement of  a single, dominant,  correlator from N-body simulations for the UV ones. We measure this correlator for a suite of seven different cosmologies, and we show that its inclusion in our perturbation scheme reproduces the fully non-linear PS with  percent level accuracy, for wave numbers up to $k\sim 0.4\, \hmpc$ down to $z=0$. 
We then show that, once this correlator has been measured in a given cosmology, there is no need to run a new simulation for a different cosmology in the suite. Indeed, by rescaling this correlator by a proper function computable in SPT, the reconstruction procedure works also for the other cosmologies and for all redshifts, with comparable accuracy. 
Finally, we clarify the relation of this approach to the Effective Field Theory methods recently proposed in the LSS context.
\end{abstract}

\maketitle

\section{Introduction}
The description of the LSS at the percent level precision is the goal of the theory of cosmological perturbations, in view of the present and next generation large galaxy surveys. N-body simulations provide a well tested tool to achieve such goal, at least in the standard $\Lambda$CDM framework. However, simulations at the required accuracy in the full range of scales, in particular including the BAO ones (with wave numbers in the $k\sim 0.05 - 0.3\;\hmpc$ range)  require large volumes and many realizations to reduce sample variance. A thorough exploration of the space of cosmological parameters and of different cosmological models is then practically impossible with these tools. 

A fast and flexible alternative is given by SPT  \cite{PT}, in particular by the improved expansion schemes developed in the recent years (see for instance \cite{Anselmi:2012cn, Bernardeau:2011dp,Taruya:2012ut}, for a recent review, see \cite{Bernardeau:2013oda}), which can reproduce the nonlinear power spectrum (PS) at the percent level in the full BAO region (and beyond \cite{Anselmi:2012cn}) down to $z=0$. However these methods have some fundamental limitations in accessing the UV scales. First of all, the nonlinearities in this regime become so pronounced that the SPT expansion does not converge any more \cite{RPTa} (for a recent analysis see \cite{Blas:2013aba}). This problem may be solved, or at least pushed to smaller scales by the improved expansions, which however, when truncated at a finite order, in general do not respect the generalized Galilean invariance of the system (for an exception, see \cite{Anselmi:2012cn, Peloso:2013zw}). Moreover, even if the Euler-Poisson system of equations from which the SPT expansion and its improvements are generated were solved exactly, it would still miss the effect of multistreaming, or small scale velocity dispersion, which is set to zero to truncate the hierarchy of equations for the moments of the distribution function, in what is known as the ``Single Stream Approximation'' (SSA) \cite{PT}. Although all these effects are mostly negligible at large scales and high redshift a precise estimation of their sizes is still lacking (for an early attempt in this direction, see \cite{Pueblas:2008uv}), and it is indeed impossible in the SPT - or, more generally, in the SSA-  framework.

In ref.~\cite{Pietroni:2011iz} an hybrid approach was proposed, which has the virtue of combining the two methods, namely N-body simulations and SPT, using each of them in the sector which it is better designed for. The idea is to introduce a smoothing scale, $L$, and to use SPT for the IR scales, defined by $k \la L^{-1}$, and N-body simulations for the UV scales, $k \ga L^{-1}$.  More concretely, a smoothing of the microscopic Vlasov equations is performed and, after taking momenta, one obtains a system of continuity, Euler and Poisson equations for the smoothed fields where now a source term appears (in the Euler equation), which encodes all the UV physics which has been smoothed out. The PS  (and higher order correlators) for the smoothed fields can now be obtained as a double expansion in initial smoothed fields (as in SPT), and in the UV source terms, which must be measured by means of N-body simulations. The advantage of this method is twofold: in the first place, the SPT expansion is in terms of smoothed fields, and is therefore much better behaved than the usual SPT expansion, which involves field fluctuations at all scales. On the other hand, N-body simulations are required only to compute the UV source terms which, being relevant only at small scales, do not require large simulation volumes. The approach was called Coarse Grained Perturbation Theory (CGPT), and, although in this work we will introduce some modifications with respect to ~\cite{Pietroni:2011iz}, we will keep the same name.

A major advantage of the approach would emerge if the UV sources would manifest some ``universal" properties, independently (or nearly independently) of the cosmological model. If this would be the case, then one would not need an independent N-body simulation for each set of cosmological parameters, and the reconstruction of the nonlinear PS by this method would become extremely fast, only requiring SPT integrals at low orders (1- or 2-loop). In this paper we investigate precisely this issue. 

First, we show that, in a given cosmology, a scheme involving just a 1-loop SPT computation and the measurement of only one cross-correlator between the source term and the smoothed density field can reproduce the nonlinear PS at the percent level for scales $k\la 0.4\;\hmpc$ down to $z=0$. The procedure can be speeded up, without spooling its accuracy, by making an Ansatz on the time dependence of this correlator (instead of measuring it directly), eq.~\re{alpha}, which results in an extra parameter to be fitted by comparing the reconstructed PS with the nonlinear one. 

Then, we fix a cosmology as the ``reference'' one and perform a set of six other N-body simulations obtained by varying in turn one out of three cosmological parameters, namely, the scalar amplitude of the fluctuations, the spectral tilt, and the matter content, by a positive and negative amount with respect to the reference cosmology. First, we repeat the reconstruction procedure for each of the new cosmologies, and we show that it works at the same level of accuracy as for the reference one. Then, we investigate the cosmological dependence of the source-density cross-correlators and find that they all have approximately the same scale dependence,  modulo a nearly scale-independent factor that can be computed by a 1-loop SPT integral. Therefore, once the correlator has been measured in a given cosmology, it can be  ``translated'' to a different one quite efficiently, without the need to measure it in a dedicated N-body simulation. 
We also show that this property holds not only at one given redshift: once the time dependence of the correlator has been measured in a given cosmology, it can be translated into any other cosmology without introducing any new free parameter to characterize it.

The approach discussed in this paper is not distant, in spirit, from the Effective Field Theory of the Large Scale Structure (EFToLSS) introduced in \cite{Baumann:2010tm,Carrasco:2012cv} and further developed in \cite{Hertzberg:2012qn,Pajer:2013jj,Carrasco:2013mua,Mercolli:2013bsa,Carroll:2013oxa,Senatore:2014via,Baldauf:2014qfa,Angulo:2014tfa}. The starting point, namely, smoothing the Vlasov equation, is indeed the same. The main difference arises in a second step, when the EFToLSS assumes an expansion of the UV source in terms of the smoothed IR fields, with some arbitrary coefficients to be fixed by fitting with the nonlinear PS, while we measure the source directly from simulations. In this paper, we also clarify the relation between these approaches.

The paper is organized as follows. In Section~\ref{method} we derive the equations for the smoothed IR fields and we define the UV sources in terms of the microscopic quantities ({\it i.e.} density, velocity, and velocity dispersion). In Section~\ref{expansionscheme} we define the expansion scheme in terms of the IR fields and of the deviation from the microscopic SSA, that is, in terms of the microscopic velocity dispersion $\sigma^{ij}$. In Section~\ref{NBODYSIM} we describe our suite of seven N-body simulations and the procedure to measure the UV source terms. In Section~\ref{PSreco} we formulate the reconstruction procedure of the nonlinear PS at 1-loop order and compare its results with the nonlinear PS from N-body simulations. In Section~\ref{cosmodep} we study the cosmology dependence of the cross-correlator between the source and the IR density field
and in Section~\ref{sec:noNbody} we use these results to perform a PS reconstruction for a cosmology without running a dedicated N-body simulation for it. In Section~\ref{sec:time} we discuss the time-dependence of the parameter describing the time-dependence of the correlator, and in Section~\ref{EFToLSS} we discuss the relation between our approach and the EFToLSS. Finally, in Section~\ref{fine} we summarize our results.

\section{Filtering the Vlasov equation}
\label{method}

The dark matter microscopic distribution function $f$ obeys the Vlasov equation  
\beq
\left(\frac{\partial}{\partial \tau}+ \frac{p^i}{ a m}\frac{\partial}{\partial x^i}- am\nabla^i_x \phi(\bx,\tau) \frac{\partial}{\partial p^i}\right)\,f(\bx,\bp,\tau)=0\,.
\label{vlasov}
\eeq

We will be interested in a smoothed version of $f(\bx,\bp,\tau)$ obtained from it by filtering with an appropriate filter function \cite{Buchert:2005xj,Baumann:2010tm,Pietroni:2011iz},
\beq
\bar f(\bx,\bp,\tau) \equiv \langle f \rangle(\bx,\bp,\tau)\,,
\eeq
where the brackets, or, equivalently, the bar, will denote the filtering operation, that is, for a generic function $g(\bx)$,
\beq
\langle g \rangle \left( \bx \right) \equiv \int d^3 y \, W \left( y/L \right) \, g \left( \bx - \by \right) \,, 
\label{filtering}
\eeq
with $W\left(x/L \right)$ a window function that has most of its support at distances~$\la L$ from $\bx$ and normalized such that 
\beq
\int d^3x \,W\left(\frac{x}{L}\right) = 1 \,.
\label{norm}
\eeq
An example of a window function with this property is a Gaussian filter
\beq
W\left(\frac{x}{L}\right)=\frac{1}{L^3}\frac{1}{(2\pi)^{3/2}}\,e^{-\frac{x^2}{2L^2}}\,. 
\eeq
This is the window function that we consider in our numerical computations. However, our analytic computations  are valid for an any function $W$ with the properties mentioned above. 
The Fourier transform of the Gaussian filter
\beq
\tilde W(k\,L)\equiv \int d^3x \,W\left(\frac{x}{L}\right)\,e^{-i \bk\cdot\bx} = e^{-\frac{k^2 L^2}{2}}\,,
\label{filterF}
\eeq
has the property, following from \re{norm}, $\tilde W(0)=1$.

Applying the filtering operation to the Vlasov equation (\ref{vlasov}),  we get
\beq
\left(\frac{\partial}{\partial \tau}+ \frac{p^i}{ a m}\frac{\partial}{\partial x^i}- am\nabla^i_x \bar\phi(\bx,\tau) \frac{\partial}{\partial p^i}\right)\,\bar f(\bx,\bp,\tau) = - K[f](\bx,\bp,\tau)\,,
\label{smoothvlasov}
\eeq
where the ``pseudo-collisional'' term generated by coarse graining is given by
\beq
\!\!\!\!\!\!\!\!\!\! \!\!\!\!\!\!\!\!\!\! \!\!\!\!\!\!\!\!\!\! K[f](\bx,\bp,\tau)\equiv - \langle am\nabla^i_x \phi(\bx,\tau) \frac{\partial}{\partial p^i}\,f(\bx,\bp,\tau)\rangle + am\nabla^i_x \bar\phi(\bx,\tau) \frac{\partial}{\partial p^i}\,\bar f(\bx,\bp,\tau)\,.
\label{mother}
\eeq
We define moments of the coarse-grained distribution function as usual,
\beqra
&& \bar n(\bx,\tau) = \int d^3 p \bar f(\bx,\bp,\tau) = \langle n \rangle(\bx,\tau)\,,\nonumber\\
&&\bar v^i(\bx,\tau)= \frac{1}{ \bar n(\bx,\tau)} \int d^3 p \frac{p^i}{am} \bar f(\bx,\bp,\tau) =  \frac{1}{1+ \bar \delta (\bx,\tau)} \langle (1+\delta) v^i \rangle(\bx,\tau)\,,\nonumber\\
&&\bar \sigma^{ij}(\bx,\tau)= \frac{1}{ \bar n(\bx,\tau)} \int d^3 p \frac{p^i}{am} \frac{p^j}{am} \bar f(\bx,\bp,\tau) - \bar v^i(\bx,\tau)\bar v^j(\bx,\tau) \nonumber\\
&&\qquad \quad\;\;\;=  \frac{1}{ 1+\bar \delta(\bx,\tau)}  \langle (1+\delta) (v^i v^j + \sigma^{ij}) \rangle- \bar v^i(\bx,\tau)\bar v^j(\bx,\tau)\,,\nonumber\\
&& \cdots\,,
\label{n-v-sigma-bar}
\eeqra
where the density fluctuations are given by 
\beq
\delta(\bx,\tau) = \frac{n(\bx,\tau)}{n_0}-1,\qquad \bar\delta(\bx,\tau) = \frac{\bar n(\bx,\tau)}{n_0}-1\,, 
\label{delta-bar}
\eeq
(with $n_0$ being the spatial average of $n \left( \bx, \tau \right)$), and  are related to the potential by the Poisson equation,
\beq
\nabla^2 \phi = \frac{3}{2}{\cal H}^2\Omega_m \,\delta\,,\qquad \nabla^2 \bar\phi = \frac{3}{2}{\cal H}^2\Omega_m \,\bar\delta\,.
\label{Poisson}
\eeq

The quantities without the bars in \re{n-v-sigma-bar}, that is $n$, $\delta$, $v^i$, $\sigma^{ij}$, are the {\em microscopic} number density, density fluctuation, velocity, and velocity dispersion, respectively, that is, the moments obtained by replacing $\bar f$ with the microscopic distribution function $f$  in the first three lines of \re{n-v-sigma-bar}.

In terms of the coarse grained fields $\bar \delta$, $\bar v^i$, and $\bar \sigma^{ij}$,  the first two moments of the Vlasov equation read  
\beqra
&&\frac{\partial}{\partial \tau} \bar \delta +\frac{\partial}{\partial x^i} \left((1+\bar \delta )\,\bar v^i  \right) =0\,, \label{continuity}\\
&&\frac{\partial}{\partial \tau} \bar v^i+{\cal H}\bar v^i + \bar v^k\frac{\partial}{\partial x^k}\bar v^i  = -\nabla_x^i\bar \phi -J^i\,,\label{euler}
\eeqra
supplemented by \re{Poisson}, where 
\beq J^i\equiv J_1^i+J_\sigma^i\,,
\label{Jtot}
\eeq with 
\beqra
&& J_1^i(\bx,\tau) = \frac{1}{1+\bar\delta(\bx,\tau)} \langle (1+\delta) \nabla^i\phi \rangle(\bx,\tau)  -  \nabla^i\bar \phi(\bx,\tau)\,, \label{J1}\\
&&J_\sigma^i(\bx,\tau) \equiv\frac{1}{1+\bar\delta} \frac{\partial}{\partial x^k}\left((1+\bar \delta) \bar \sigma^{ik}\right).
\label{Jsig}
\eeqra
Notice that $J_1^i$ is just the first moment of \re{mother} divided by $1+\bar \delta$ and that it vanishes if $\delta$ is constant inside the filtering volume.  The contribution of the next moment, $\bar\sigma^{ij}$, is fully taken into account by the source term $J_\sigma^i$, that is, the set of equations \re{continuity}, \re{euler} and the second of \re{Poisson}, is exact, although it requires external input on the sources \re{Jtot}.

The unfiltered fields obey  the same set of equations as the filtered ones, but  without source, $J^i =0$.

We decompose the coarse-grained velocity in divergence (${\bar \theta} \equiv {\bf \nabla} \cdot {\bf {\bar v}}$) plus vorticity. In Fourier space, 
\begin{equation}
{\bar v}^i \left( \bk \right) \equiv - i \frac{k^i}{k^2} {\bar \theta} \left( \bk \right) + i \, \epsilon_{ijl} \frac{k^j}{k^2} {\bar v}_*^l \left( \bk \right) ,
\label{vbar-deco}
\end{equation} 
where  $ {\bar v}_*^l \left( \bk \right) \equiv i \,\epsilon_{lmn} k^m {\bar v}^n \left( \bk \right) $ is the vorticity field (which has only two independent components, since $ k^l {\bar v}_*^l \left( \bk \right) =0 $).
Taking the divergence of the  Euler equation \re{euler}, we obtain
\begin{eqnarray}
&& \!\!\!\!\!\!\!\!\!\!\!\! \!\!\!\! \!\!\!\!\partial_\tau {\bar \theta} \left( \bk \right) + {\cal H}   {\bar \theta} \left( \bk\right) + \frac{3}{2} {\cal H}^2 \Omega_m \delta \left( \bk \right) = - i k_i J_\bk^i 
 + \frac{I_{\bk;\bp,\bq}}{p^2 q^2} \Bigg\{ - \frac{k^2 \bp \cdot \bq}{2} {\bar \theta } \left( \bp \right)  {\bar \theta } \left( \bq \right) 
\nonumber\\ 
&& \!\!\!\! \!\!\!\! \!\!\!\!
+ \left[ \bp \cdot \left(  \bp + 2 \bq \right) \right] {\bar \theta } \left( \bp \right) \left[ \left( \bp \times \bq \right) \cdot {\bar \bv}_*\left( \bq \right) \right] 
 +  \left[ \left( \bp \times \bq \right) \cdot {\bar \bv}_* \left( \bp \right) \right]   \left[ \left( \bp \times \bq \right) \cdot {\bar \bv}_* \left( \bq \right) \right]  \Bigg\} , \nonumber\\ 
 \label{eulerbar}
\end{eqnarray} 
where we have introduced  the symbol
\beq
I_{\bk;\bq_1,\cdots \bq_n}\equiv \int \frac{d^3 q_1}{(2\pi)^3} \cdots  \frac{d^3 q_n}{(2\pi)^3} (2\pi)^3\delta_D(\bk-\sum_{i=1}^n \bq_i)\,.\
\eeq

We will treat ${\bar \theta}$ as a dynamical field in our computations. 
We instead treat the vorticity as a source term (analogously to $J_1^i $ and $J_\sigma^i$). As we discuss below, we will be only interested in a perturbative expansion of it. Starting from
\begin{equation}
{\bar v}^i = \langle v^i \rangle + \langle \delta v^i \rangle - {\bar \delta} \langle v^i \rangle + {\rm O } \left( \delta^2 v \right) \;\;,\;\; {\bar \theta} = \partial_i {\bar v}^i \;, 
\end{equation} 
and assuming that the vorticity only originates from the filtering procedure   (namely, that the microscopic velocity has zero vorticity, $v^i = - i \frac{k^i}{k^2 } \theta$, as it is assumed in PT), we can express it in terms of the density and velocity divergence fields, as 
\begin{eqnarray}
\label{vstar-pert}
\!\!\!\!\!\!\!\!\!\!\!\!\!\!\!\! i \epsilon_{ijl} \frac{k_j}{k^2} v_*^l &\equiv& {\bar v}^i + \frac{i k^i}{k^2} {\bar \theta} \nonumber \\ 
&  &\!\!\!\!\!\!\!\! \!\!\!\!\!\!\!\!\!\!\!\!\!\!\!\! = - i I_{\bk;\bp,\bq} \left( \frac{q^i}{q^2} - \frac{k^i \,  k \cdot q}{k^2 q^2} \right) \Big( \left\langle \delta \left( \bp \right) \theta \left( \bq \right) \right\rangle 
- {\bar \delta} \left( \bp \right) \left\langle \theta \left( \bq \right) \right\rangle \Big) + {\rm O } \left( \delta^2 v \right).
\end{eqnarray}
An alternative method would be to measure the vorticity field from N-body simulations and to use it as an extra source term of the field equations,  as it was done in  \cite{Pueblas:2008uv}, where it was shown that the vorticity contribution to the PS is at the sub percent level in the whole BAO range of scales at $z=0$.

\subsection{Compact notation}

It is convenient to introduce the time variable $\eta \equiv  \ln \frac{D_+ \left( \tau \right)}{D_+ \left( \tau_{\rm in} \right)}$, where $D_+$ is the linear growth factor and $\tau_{in}$ an initial time chosen in such a way that all the relevant scales are still in the linear regime.  We define the fields
\begin{equation}
\!\!\!\!\!\!\!\!\! \!\!\!\!\!\!\!\!\!  {\bar \vp}_1 \left( \eta , \bk \right) \equiv {\rm e}^{-\eta} {\bar \delta } \left( \eta , \bk \right) \;\;,\;\;
{\bar \vp}_2 \left( \eta , \bk \right) \equiv  {\rm e}^{-\eta}  \frac{-{\bar \theta } \left( \eta , \bk \right) }{{\cal H } f} \;\;,\;\;
{\bar \omega}^i \left( \eta , \bk \right) \equiv \frac{{\bar v}_*^i \left( \eta , \bk \right)}{{\cal H} f} , 
\label{phi1-2}
\end{equation}
where we have also used the linear growth function $f \left( \eta \right) \equiv \frac{1}{\cal H} \frac{d \eta}{d \tau}$. 
In terms of these fields, the two dynamical equations for the filtered density (eq.~\re{continuity}) and velocity divergence (eq.~\re{eulerbar}) (supplemented by the Poisson equation \re{Poisson}), acquire the compact form 
\beqra
 &&\!\!\!\!\!\!\!\!\! \!\!\!\!\!\!\!\!\! \!\!\!\!\!\!\!\!\!(\delta_{ab}\partial_\eta +\Omega_{ab})\vpb_b(\bk,\eta)= I_{\bk,\bq_1,\bq_2}  e^\eta\gamma_{abc}(\bq_1,\bq_2) \vpb_b(\bq_1,\eta)\vpb_c(\bq_2,\eta) -h_a(\bk,\eta)  \nonumber\\
&&\;\;\;\; \qquad\qquad-h_a^\omega(\bk,\eta)+ I_{\bk,\bq_1,\bq_2}  A_a^j(\bq_1,\bq_2)\vpb_a(\bq1,\eta)\bar\omega^j(\bq_2,\eta)\,.
\label{eom}
\eeqra
The LHS of this relation is the linearized part of the equation for the two dynamical modes, and it is characterized by 
\begin{equation}
{\bf \Omega} = \left( \begin{array}{cc} 1 & - 1 \\ - \frac{3}{2} \frac{\Omega_m}{f^2} &  \frac{3}{2} \frac{\Omega_m}{f^2} \end{array} \right) .
\end{equation}
The first term at the RHS of eq. (\ref{eom}) encodes the interaction between the dynamical fields. The only nonvanishing components of the vertex functions are 
\begin{equation}
\!\!\!\!\!  \!\!\!\!\!  \!\!\!\!\!  \!\!\!\!\!  \!\!\!\!\!    \gamma_{121} \left( \bp , \bq \right) = \frac{\left( \bp+\bq \right) \cdot \bp}{2 p^2} \;\;,\;\;
\gamma_{112} \left( \bq , \bp \right) = \gamma_{121} \left( \bp , \bq \right) \;\;,\;\; 
\gamma_{222} = \frac{\left( \bp + \bq \right)^2 \bp \cdot \bq}{2 p^2 q^2} \,.
\label{gamma}
\end{equation} 
the second term on the RHS of  eq. (\ref{eom}) is the contribution from the source term in eq.~\re{Jtot}:
\beq
h_a(\bk,\eta) \equiv h_a^1(\bk,\eta)+h_a^\sigma(\bk,\eta)\,,
\label{htot}
\eeq
where
\beq
\!\!\!\!\!  \!\!\!\!\! \!\!\!\!\!    h_a^1(\bk,\eta)=-i\,\frac{k^i J_1^i(\bk,\eta)}{{\cal H}^2 f^2}e^{-\eta}\, \delta_{a2}\,,\qquad\qquad  h_a^\sigma(\bk,\eta)=-i\,\frac{k^i J_\sigma^i(\bk,\eta)}{{\cal H}^2 f^2}e^{-\eta} \,\delta_{a2}\,.
 \label{hsum}
\eeq
 In the last two terms  on the RHS of    eq. (\ref{eom}) we have instead 
\beqra
&&h_a^\omega(\bk,\eta)= I_{\bk;\bq_1,\bq_2} \frac{(\bq_1\times\bq_2)^i(\bq_1\times\bq_2)^j}{q_1^2 q_2^2} \bar\omega^i(\bq_1,\eta)\bar\omega^j(\bq_2,\eta) {\rm e}^{-\eta}  \,\delta_{a2}\,, \nonumber\\ 
&&A_1^j(\bq_1,\bq_2) =\frac{(\bq_1\times\bq_2)^j}{q_2^2} \,,\quad A_2^j(\bq_1,\bq_2)=\frac{q_1^2+2\bq_1\cdot\bq_2}{q_1^2}   A_1^j(\bq_1,\bq_2)\,.
\label{sources_vort}
\eeqra

\section{Expansion scheme }
\label{expansionscheme}

We will split the source term as
\beq
h_a(\bk,\eta)=h_{ss, a}(\bk,\eta)+\delta h_a(\bk,\eta)\,,
\label{sourcesplit}
\eeq
where $h_{ss,a}(\bk,\eta)$ is the source term induced by the filtering procedure in the SSA recalled in the Introduction, namely, in the approximation in which the microscopic velocity dispersion $\sigma^{ij}$ and all the higher order moments vanish\footnote{These microscopic moments can be thought as being measured in a N-body simulation of infinite resolution, in the limit $L\to 0$. }.  More precisely, $h_{ss,a}(\bk,\eta)$ is given by
\beq
h_{ss,a}(\bk,\eta) \equiv h_{ss,a}^1(\bk,\eta)+h_{ss,a}^\sigma(\bk,\eta)\,,
\label{htotss}
\eeq
where
\beq
\!\!\!\!\!  \!\!\!\!\!  \!\!\!\!\! \!\!\!\!\!    h_{ss,a}^1(\bk,\eta)=-i\,\frac{k^i J_{1,ss}^i(\bk,\eta)}{{\cal H}^2 f^2}e^{-\eta} \,\delta_{a2}\,,\qquad  h_{ss,a}^\sigma(\bk,\eta)=-i\,\frac{k^i J_{\sigma,ss}^i(\bk,\eta)}{{\cal H}^2 f^2}e^{-\eta} \,\delta_{a2}\,,
 \label{hsumss}
\eeq
with
\beqra
&&J_{1,ss}^i(\bx,\tau) = \frac{1}{1+\bar\delta_{ss}(\bx,\tau)} \langle (1+\delta_{ss}) \nabla^i\phi_{ss} \rangle(\bx,\tau)  -  \nabla^i\bar \phi_{ss}(\bx,\tau)\,, \nonumber\\
&&J_{\sigma,ss}^i(\bx,\tau) \equiv\frac{1}{1+\bar\delta_{ss}} \frac{\partial}{\partial x^k}\left((1+\bar \delta_{ss}) \bar \sigma_{ss}^{ik}\right), 
\label{J1J2ss}
\eeqra
where the velocity dispersion is generated only by the filtering procedure,
\beqra
\bar \sigma_{ss}^{ij}(\bx,\tau)=  \frac{1}{ 1+\bar \delta_{ss}(\bx,\tau)}  \langle (1+\delta_{ss}) (v^i_{ss} v^j_{ss}) \rangle- \bar v^i_{ss}(\bx,\tau)\bar v^j_{ss}(\bx,\tau)\,,
\eeqra
and the microscopic $\delta_{ss}$ and $v^i_{ss}$ fields, grouped as usual in $\vp_{ss,a}$, obey the field equations in the single stream approximation
\beqra
&&(\delta_{ab}\partial_\eta +\Omega_{ab})\vp_{ss,b}(\bk,\eta)= I_{\bk,\bq_1,\bq_2}  e^\eta\gamma_{abc}(\bq_1,\bq_2) \vp_{ss,b}(\bq_1,\eta)\vp_{ss,c}(\bq_2,\eta)\,.
\label{eomss}
\eeqra
The filtered fields in the single stream approximation, $\vpb_{ss,a}$, on the other hand, satisfy eq.~\re{eom} with $\delta h_a(\bk,\eta)=0$, that is, with $h_a(\bk,\eta)=h_{ss, a}(\bk,\eta)$. In this equation, the vorticity induced by the filtering procedure will also be kept, for consistency.\footnote{ Specifically, we refer to the terms \re{sources_vort}.  Such terms are known to be  subdominant contributions \cite{Pueblas:2008uv}. However they are needed to  check perturbatively the cutoff-independence of the results in an expansion series in the linear fields $\vp^{lin}$. Therefore we retain them when we perform this expansion (see below). }

The full solution for the filtered field $\vpb_a(\bk,\eta)$ can then be formally written as an expansion in the microscopic velocity dispersion $\sigma^{ij}$,
\beq
\vpb_a(\bk,\eta) = \vpb_{ss,a}(\bk,\eta) + \sum_{m=1}^{\infty} \vpb_{\sigma,a}^{(m)}(\bk,\eta)\,, 
\label{expvpb}
\eeq
where $ \vpb_{\sigma,a}^{(m)}(\bk,\eta) $ is of ${\rm O } \left( \sigma^m \right)$. Analogously, we expand the unfiltered field as 
\beq
\vp_a(\bk,\eta) = \vp_{ss,a}(\bk,\eta) + \sum_{m=1}^{\infty} \vp_{\sigma,a}^{(m)}(\bk,\eta)\,.
\eeq
We note that,  if we consider the density field (namely, $a=1$), we have, order by order in $\sigma$, 
\beq
 \vpb_{\sigma,1}^{(m)}(\bk,\eta) = \tilde W(k) \, \vp_{\sigma,1}^{(m)}(\bk,\eta)\,,
 \label{filts}
\eeq
besides the single stream relation $ \vpb_{ss,1}(\bk,\eta) =\tilde W(k)\, \vp_{ss,1}(\bk,\eta) $. This is because the filtering procedure commutes with the expansion in the microscopic dispersion. 

For bookkeeping, it is also useful to expand the source 
\beq
\delta h_a(\bk,\eta) = \sum_{m=1}^{\infty} \delta h_{\sigma,a}^{(m)}(\bk,\eta)\,.
\eeq

Using eq.~\re{expvpb} we can compute the contributions to the PS  from the correlator 
\beqra
&&\!\!\! \!\!\! \!\!\! \langle \vpb_a(\bk,\eta) \vpb_b(\bk',\eta') \rangle = \langle \vpb_{ss,a}(\bk,\eta) \vpb_{ss,b}(\bk',\eta') \rangle \nonumber\\
 &&\;\;\;\;\;  + \langle \vpb_{\sigma,a}^{(1)}(\bk,\eta) \vpb_{ss,b}(\bk',\eta') \rangle +
 \langle \vpb_{ss,a}(\bk,\eta) \vpb_{\sigma,b}^{(1)}(\bk',\eta') \rangle +O(\langle \sigma^2\rangle)\,.
 \label{linexp}
\eeqra
Using  eq. (\ref{eom}), the $\vpb_{\sigma,a}^{(1)}$ terms at the last line of eq. \re{linexp}, can be written as 
\begin{eqnarray}
\vpb_{\sigma,a}^{(1)}(\bk,\eta) &=& - \int_{\eta_{\rm in}}^\eta d s \, g_{ac} \left( \eta - s \right) \delta h_{c}^{(1)} \left( \bk, s \right) + {\rm O } \left(   \sigma \vp^{lin} \right)  \nonumber\\ 
 &=& - \int_{\eta_{\rm in}}^\eta d s \, g_{ac} \left( \eta - s \right) \left[  h_{c} \left( \bk, s \right) - h_{ss,c}  \left( \bk, s \right) \right]  + {\rm O } \left( \sigma^2 , \sigma \vp^{lin} \right) \,,  \nonumber\\ 
\label{phisig1}
\end{eqnarray} 
where the propagator $g_{ab}$ is the inverse of the operator at LHS of (\ref{eom}): 
\begin{equation}
\left( \delta_{ab} \partial_\eta - \Omega_{ab} \right) g_{bc} \left( \eta - s \right) = \delta_{ac} \delta_D \left( \eta - s \right) 
\end{equation} 
The terms at the last line of eq. \re{linexp} are therefore \footnote{We note that we are careful in keeping track of the correct $ \vp^{lin} $-dependence of the ${\rm O } \left( \sigma \right)$ correction, while we simply  denote the terms proportional to the second power of the microscopic velocity dispersion as ${\rm O } \left( \sigma^2 \right)$. } 
\begin{eqnarray}
&& \!\!\!\!\!\!\!\! \!\!\!\!\!\!\!\!  \!\!\!\!\!\!\!\! \!\!\!\!\!\!\!\!  \!\!\!\!\!\!\!\!  \!\!\!\!\!\!\!\! 
\left\langle \vpb_{\sigma,a}^{(1)}(\bk,\eta) \vpb_{ss,b}(\bk',\eta') \right\rangle =  \nonumber\\
&&  \!\!\!\!\!\!\!\! \!\!\!\!\!\!\!\!  \!\!\!\!\!\!\!\! \!\!\!\!\!\!\!\!  \!\!\!\!\!\!\!\!  \!\!\!\!\!\!\!\! 
 - \int_{\eta_{in}}^\eta ds \; g_{ac}(\eta-s) \left\langle 
 \left[  h_{c} \left( \bk, s \right) - h_{ss,c}  \left( \bk, s \right) \right]    \vpb_{ss,b}(\bk',\eta') \right\rangle 
 + {\rm O } \left( \left\langle \sigma^2\,\vp^{lin} , \sigma \left( \vp^{lin} \right)^2 \right\rangle \right)  \nonumber\\
&&  \!\!\!\!\!\!\!\! \!\!\!\!\!\!\!\!  \!\!\!\!\!\!\!\! \!\!\!\!\!\!\!\!  \!\!\!\!\!\!\!\!  \!\!\!\!\!\!\!\! 
= - \int_{\eta_{in}}^\eta ds \; g_{ac}(\eta-s) \Bigg[ \left\langle 
 h_{c} \left( \bk, s \right) \vpb_{b}(\bk',\eta') \right\rangle - \left\langle  h_{ss,c}  \left( \bk, s \right) \vpb_{ss,b}(\bk',\eta')    \right\rangle\Bigg] 
 + {\rm O } \left( \left\langle \sigma^2, \sigma \left( \vp^{lin} \right)^2 \right\rangle \right) \,, \nonumber\\ 
 \label{app1}
\end{eqnarray}
where we have replaced $ \vpb_{ss,b}$ with $  \vpb_{b}$ in the first correlator, to get a term which is directly measurable from simulations (the difference between the two expressions gives a contribution of $ {\rm O } \left( \sigma^2 \right)$, which we are   already disregarding in (\ref{linexp})).  Inserting   \re{app1} in \re{linexp}, gives
\beqra
&&\!\!\! \!\!\! \!\!\! \langle \vpb_a(\bk,\eta) \vpb_b(\bk',\eta') \rangle = \langle \vpb_{ss,a}(\bk,\eta) \vpb_{ss,b}(\bk',\eta') \rangle \nonumber\\
 && +  \int_{\eta_{in}}^\eta ds \; g_{ac}(\eta-s) \Bigg[ \left\langle 
 h_{ss,c}  \left( \bk, s \right)    \vpb_{ss,b}(\bk',\eta') \right\rangle- \left\langle 
 h_{c}  \left( \bk, s \right)    \vpb_{b}(\bk',\eta') \right\rangle\nonumber\\
 &&\qquad\qquad\qquad \qquad\qquad\quad +\bigg(a \leftrightarrow b, \bk\leftrightarrow \bk' \bigg)\Bigg] +O(\sigma^2 , \sigma (\vp^{lin})^2)\,.
 \label{rec1}
\eeqra

We rewrite this expression (taking $\eta' = \eta$) as 
\beqra
&& \!\!\!\!\!\!\!\!  \!\!\!\!\!\!\!\!  \!\!\!\!\!\!\!\!  \!\!\!\!\!\!\!\! 
\bar P_{ab} (k,\eta) =  \bar P_{ss,ab} 
(k,\eta) - \Delta \bar P_{ss,ab}^h (k,\eta) \nonumber\\
&& \!\!\!\!\!\!\!\!  \!\!\!\!\!\!\!\!  - \left( \int_{\eta_{in}}^\eta ds \; g_{a c}(\eta-s)  \left\langle 
 h_{c}  \left( \bk, s \right)    \vpb_{b}(\bk',\eta') \right\rangle'  
 +\left( a \leftrightarrow b, \bk\leftrightarrow \bk' \right)  \right)  
  +{\rm O}(\sigma^2 , \sigma (\vp^{lin})^2
   )\,, \nonumber\\ 
\label{rec2}
\eeqra
where prime over an expectation value denotes the expression without the corresponding $\delta_D$-function: 
\begin{equation}
\langle f \left( \bk \right) g \left( \bk' \right) \rangle \equiv \left( 2 \pi \right)^3 \delta_D \left( \bk + \bk' \right) \langle f \left( \bk \right) g \left( \bk' \right) \rangle' \,. 
\end{equation}

The meaning of eq.~\re{rec2} is straightforward: the PS is obtained from the {\em fully nonlinear PS in the single stream approximation} (the first term at RHS), to which we subtract the correlator which includes the short-distance nonlinearities in the single stream approximation (the second term), and add back the correlator which includes both the short-distance nonlinearities and the deviations from the single stream approximation (the third term). 

At the order indicated in the equation, the dependence on the filter function of the RHS  equals that of the LHS. In particular, it gives a simple $\tilde W[k L]^2$ factor for the density-density correlator ($a=b=1$). Besides this trivial dependence, there is no extra dependence on the value of the $L$-cutoff. In practice, however, the single stream contributions to eq.~\re{rec2} must be computed in some approximation, which will inevitably induce a spurious dependence on the filtering procedure. Computing the first two contributions to eq.~\re{rec2} at $n$-th loop order induces a spurious cutoff-dependence at the $n+1$-th loop order. The value of the cutoff in this case has to be chosen in such a way that a PT computation at the considered finite order is reliable.  Therefore, the value of the cutoff $L$ will be chosen large enough that a 1-loop computation is reliable for the large scales ($k\la1/L$) (so to minimize the ${\rm O} \left( (\vp^{lin})^6 \right)$ corrections),  and, at the same time small enough so that  the $O(\sigma^2 , \sigma (\vp^{lin})^2)$ contributions that we neglect are indeed subdominant.  The value $L=2 \; \mpch$ has been used in in our numerical computations. We discuss this choice in \ref{choice-L}. 

In this paper, we will limit the PT computation of the large distance (single stream) scales at the 1-loop order, giving rise to 
\beqra
&& \!\!\!\!\!\!\!\!  \!\!\!\!\!\!\!\! 
\bar P_{ab} (k,\eta) =  \bar P_{ss,ab} ^{lin}(k,\eta)+  \bar P_{ss,ab} ^{\rm 1-loop}
(k,\eta) -  \Delta \bar P_{ss,ab}^{h,{\rm 1-loop}} (k,\eta) \nonumber\\
&&\quad - \left( \int_{\eta_{in}}^\eta ds \; g_{a 2}(\eta-s)  \left\langle 
 h_{2}  \left( \bk, s \right)    \vpb_{b}(\bk',\eta') \right\rangle'  
 +\left( a \leftrightarrow b, \bk\leftrightarrow \bk' \right)  \right)  
  \nonumber\\
&& \qquad\qquad 
  +{\rm O}(\sigma^2 , \sigma (\vp^{lin})^2,  (\vp^{lin})^6
   )\,. 
 \label{Pr1l}
\eeqra
 The 1-loop quantities, $ \bar P_{ss,11}^{\rm 1-loop} $, and  $ \Delta \bar P_{ss,11}^{h,{\rm 1-loop}}  $ are computed in   \ref{check-factorization}, and their final expressions are given in \ref{app-explicit}. 
 In the next section we will describe the procedure for extracting the correlators at the second line of \re{Pr1l} from N-body simulations.

\section{N-body simulations and measurement of the sources}
\label{NBODYSIM}

We have run large box-size N-body simulations using the TreePM code
{\sc GADGET-II} \cite{Springel:2005mi}. Our simulations follow the
evolution, until $z=0$, of $n_{part}=512^3$ cold dark matter (CDM) particles
within a periodic box of  $L_{box} = 512 \;\mpch $  comoving. The initial
conditions were generated at $z=99$ by displacing the positions of the
CDM particles, that were initially set in a regular cubic grid, using
the Zel'dovich approximation. The Coyote simulations we compare our results with 
\cite{Lawrence:2009uk,Heitmann:2013bra} also uses the Zel'dovich approximation for initial conditions, at the initial redshift $z_{\rm in} = 200$. As discussed in \cite{Heitmann:2008eq}  (see their figure A19), we do not expect that the difference between their and our $z_{\rm in}$ has a substantial impact on our findings. We have also run a simulation starting at z=99 but generating the ICs with 2LPT, and we found that in terms of the matter power spectrum the differences with respect to our fiducial model are below the percent level for the scales studied in this paper, confirming previous results \cite{Crocce:2006ve}.  The difference for the $\left \langle h_2  \vpb_{1} \right\rangle'$ correlator was below $1.3\%$ at $z=0$.  This correlator contributes to the total PS only at the $\sim 10\%$ level
(as can be seen from Figure \ref{fig:rec-alpha}), and therefore our results are not affected by this difference. To give a quantitative comparison: in the first line of Table \ref{tab:REF} we obtained the value $\alpha_{\rm rec.} = 1.76_{-0.05}^{+0.06}$ 
with an error ${\rm err.}_{\rm rec.} = 1.0\%$. Giving the  ICs with 2LPT we obtained  $\alpha_{\rm rec.} = 1.75_{-0.05}^{+0.05}$ and the same ${\rm err.}_{\rm rec.}$. 

The transfer functions and the matter
power spectra have been computed at the initial redshift with CAMB
\cite{CAMB}. Baryonic effects in the initial linear power spectrum are
accounted for by using a transfer function that is a mass-weighted
average of the CDM and baryon transfer functions. The Plummer-equivalent gravitational softening of the particles is set to 60 $h^{-1}$ com. kpc. In our simulations, the  number of cells in the PM algorithm (particle-mesh) is $N=512^3 $ (this is the PMGRID parameter  \cite{Springel:2005mi}), and we have explicitly verified that our results do not change if we vary this number.   Specifically, changing the number of cells in the PM algorithm  to   $N=1024^3$, the first line of Table \ref{tab:REF} gets modified into  $\alpha_{\rm rec.} = 1.75_{-0.05}^{+0.06}$ and no change in ${\rm err.}_{\rm rec.}$ (the  $\left \langle h_2  \vpb_{1} \right\rangle'$ correlator is modified below the $\%$ level at $z=0$  at the scales of our interest).

A summary of the simulation suite is shown in Table~\ref{tab_sims}.

\begin{table}
\begin{center}
\begin{tabular}{|c|c|c|c|c|c|c|}
\hline
Name  & $\Omega_{\rm m}$ & $\Omega_{\rm b}$ & $\Omega_\Lambda$ & h & $n_s$ & $A_s \left[ 10^{-9} \right] $ \\
\hline 
REF  & 0.271 & 0.045 & 0.729 & 0.703 & 0.966 & 2.42 \\ 
\hline
$A_s^-$  & 0.271 & 0.045 & 0.729 & 0.703 & 0.966 & 1.95\\ 
\hline
$A_s^+$  & 0.271 & 0.045 & 0.729 & 0.703 & 0.966 & 3.0\\ 
\hline
$n_s^-$  & 0.271 & 0.045 & 0.729 & 0.703 & 0.932 &  2.42 \\ 
\hline
$n_s^+$  & 0.271 & 0.045 & 0.729 & 0.703 & 1.000 &  2.42 \\ 
\hline
$\Omega_{\rm m}^-$  & 0.247 & 0.045 & 0.753 & 0.703 & 0.966 &  2.42 \\ 
\hline
$\Omega_{\rm m}^+$  &  0.289 & 0.045 & 0.711 & 0.703 & 0.966 &  2.42 \\ 
\hline
\end{tabular}
\end{center} 
\caption{Summary of the simulation suite. All simulations contain $512^3$ CDM particles, and have a comoving box size of $512 \, {\rm h^{-1} \,Mpc}$.  The simulation named REF is our reference model simulation.  The name of the other simulations indicates the parameter that varies with respect to the reference model and the superscript denotes whether that value is higher ($+$) or lower ($-$) than the corresponding one of the reference model.}
\label{tab_sims}
\end{table}

We compute the value of the density contrast $\delta$, the velocity $v^i$, the acceleration $a^i$ and the velocity dispersion $\sigma^{ij}$ in each of the $N^3 = 256^3$ points placed on a regular cubic grid, covering the whole 
simulation volume using the Cloud-In-Cell (CIC) interpolation procedure
\footnote{Notice that the CIC interpolation is equivalent to trilinear interpolation.}. We now detail the procedure used
to compute them. The value of the density contrast in any point of the grid,
$\vec{r_g}=(x_g,y_g,z_g)$, is obtained by calculating:
\beq
\delta(\vec{r}_g)=\frac{N^3}{n_{\rm part} } \, 
\sum_{i=1}^{n_{\rm part}}  {\cal {\cal W}}(x_g-x_i){\cal W}(y_g-y_i){\cal W}(z_g-z_i) - 1 \;, 
\label{CIC-n}
\eeq
where  $\vec{r_i}=(x_i,y_i,z_i)$ is the location of the $i-$th particle, and  ${\cal W}(x)$ is the CIC window function 
\beq
{\cal W}(x) = \left\{ 
  \begin{array}{l l}	
    1 - \frac{ \vert x \vert}{d} & \quad {\rm if} ~ \vert x \vert \textless d
    \\
    0 & \quad {\rm otherwise}\\
  \end{array} \right. \,, 
\eeq
where $d=L_{box}/N=2\;\mpch$ is the cell size. Similarly, the value of any velocity or acceleration along a given axis, for a particular grid point, is computed as:
\beq
A(\vec{r}_g)=\frac{\sum_{i=1}^{n_{\rm part}} A_i \, {\cal W}(x_g-x_i){\cal W}(y_g-y_i){\cal W}(z_g-z_i)}{\sum_{i=1}^{n_{\rm part}}  {\cal W}(x_g-x_i){\cal W}(y_g-y_i){\cal W}(z_g-z_i)} \,, 
\label{CIC-va}
\eeq
with $A_i$ being a velocity or acceleration component of the i-th particle. Finally, the values of $\sigma^{jk}$ are calculated as:
\beq
\!\!\!\!\!\!\! \!\!\!\!\!\!\!  \!\!\!\!\!\!\! \sigma^{jk}(\vec{r}_g)=
\frac{\sum_{i=1}^{n_{\rm part}} (v_{i}^j-v^j{(\vec{r}_g}))(v_{i}^k-v^k{(\vec{r}_g})) {\cal W}(x_g-x_i){\cal W}(y_g-y_i){\cal W}(z_g-z_i) 
}{\sum_{i=1}^{n_{\rm part}}  {\cal W}(x_g-x_i){\cal W}(y_g-y_i){\cal W}(z_g-z_i)} \,, 
\label{CIC-s}
\eeq
with $v_{i}^j$ being the velocity of the particle $i$ along axis $j$ and $v^j(\vec{r}_g)$ the velocity field, along axis $j$, in the grid point $\vec{r}_g$ computed as above.
We use the N-body data to evaluate   the sources $h^1$ and $h^\sigma$; up to an overall factor, they are given by the divergences of the currents $\vec{J}_1$ and $\vec{J}_\sigma$,  cf.  eq. \re{hsum}. To compute the two currents, we first start from the quantities obtained in \re{CIC-n}, \re{CIC-va}, and \re{CIC-s}, and  we compute ${\bar n}, {\bar v}^j, {\bar \sigma}^{jk}$ as indicated by eq.  \re{n-v-sigma-bar}, after using the filter function, in Fourier space, given in eq.~\re{filterF}. We then compute the current $J_1^i$ according to \re{J1}, where the gravitational potential is obtained from the acceleration field and $\frac{1}{a} \nabla^i {\phi} = -  a^i$ (where $a$ at the LHS is the scale factor - this is necessary because the gradient in this relation is a comoving gradient). The current $J_\sigma^i$ is instead computed according to \re{Jsig}.

\section{PS reconstruction}
\label{PSreco}

Setting $a=b=1$ in eq. \re{Pr1l} gives our expression for the PS of the density field: 
\beq
\!\!\!\!\!  \!\!\!\!\!  \!\!\!\!\!  \!\!\!\!\!  \bar P_{11} (k,\eta) \simeq  \bar P_{11} ^{lin}
(k,\eta)+  \bar P_{ss,11} ^{\rm 1-loop}
(k,\eta)  -  \Delta \bar P_{ss,11}^{h,{\rm 1-loop}} (k,\eta)+  \Delta \bar P_{11}^{h,{\rm N-body}} (k,\eta) \,, 
\label{P-reconstructed} 
\eeq
where 
\beq
 \Delta \bar P_{11}^{h,{\rm N-body}} (k,\eta) \equiv  - 2  \int_{\eta_{in}}^\eta ds \; g_{1 2}(\eta-s)  \left\langle 
 h_2  \left( \bk, s \right)    \vpb_{1}(\bk',\eta) \right\rangle'  \,,
 \label{ccint}
\eeq
and we recall that the explicit expressions for the 1-loop PS are given in   \ref{app-explicit}. 

To evaluate this expression, we would  need to compute the unequal time correlator between the source $h_2 \left(\bk, s \right)$ and the filtered field $\vpb_1 \left( \bk, \eta \right)$, and integrate over the time $s$. To perform this computation, we  would need to record and cross-correlate a large number of simulation snapshots. To get a faster procedure, we assume that the correlator has a simple power law dependence on the linear growth factor,
\begin{equation}
\left \langle h_2 \left( \bk, s \right)   \vpb_{1}(\bk',\eta) \right\rangle'  \cong \left[ \frac{D \left( s \right)}{D \left( \eta \right) } \right]^{\alpha \left( \eta \right) } \times \left \langle h_2 \left( \bk, \eta \right)   \vpb_{1}(\bk',\eta) \right\rangle'   \;\;,\;\; s < \eta \;.
\label{alpha}
\end{equation} 
Computing the correlator in lowest order (1-loop) SPT would give 
\beq
\alpha \left( \eta \right) = \alpha_{SPT} = 2\,,
\eeq
which should be valid at high redshift, as we will verify later. At late times we expect a deviation from $\alpha_{PT}$, which is due to UV effects beyond 1-loop SPT.

Inserting the Ansatz \re{alpha} into eq. \re{P-reconstructed} and using the explicit form of the linear propagator, the time integration can be performed analytically, giving 
\beq
\Delta \bar P_{11}^{h,{\rm N-body}} (k,\eta) \simeq  - \frac{4}{\alpha \left( \eta \right) \left[ 5 + 2 \alpha \left( \eta \right) \right] }  \left\langle 
 h_2  \left( \bk, \eta \right)    \vpb_{1}(\bk',\eta) \right\rangle'  \,.  
 \label{P-reconstructed2} 
\eeq

We treat $\alpha \left( \eta \right)$ as a free parameter, that we obtain by minimizing the difference between the LHS and the RHS of eq. \re{P-reconstructed}. As we show below (see also  Section \ref{sec:time})   the values of $\alpha$ obtained with this method are consistent with those that can be inferred from the time dependence of the correlators, thus corroborating our Ansatz for the time dependence, eq.~\re{alpha}.

To perform the minimization, we define the ratio between the RHS and the LHS of eq. \re{P-reconstructed}, 
\beqra
&&\!\!\!\!\! \!\!\!\!\!  \!\!\!\!\!  \!\!\!\!\!  \!\!\!\!\!  \!\!\!\!\! \!\!\!\!\!    R \left[ \alpha \left( \eta \right) ; k, \eta \right] \equiv \nonumber\\
&&\!\! \!\!\! \!\!\!\!\! \!\!\!\!\!  \!\!\!\!\!  \!\!\!\!\! \!\!\!\!\!     \frac{  \bar P_{11} ^{lin} (k,\eta) + \bar P_{ss,11} ^{\rm 1-loop} (k,\eta) -  \Delta \bar P_{ss,11}^{h,{\rm 1-loop}} (k,\eta) 
    - \frac{4}{\alpha \left( \eta \right) \left[ 5 + 2 \alpha \left( \eta \right) \right] }  \left\langle 
 h_2  \left( \bk, \eta \right)    \vpb_{1}(\bk',\eta) \right\rangle'  }{\bar P_{11} (k,\eta) }  \,,
\label{ratio-R}
 \eeqra
and the average relative error 
\begin{equation}
{\cal E} \left[ \alpha \left( \eta \right) ;  \eta \right] \equiv  \sqrt{  \frac{1}{0.3 \, \hmpc} \, \int_{0.1 \, \hmpc}^{0.4 \, \hmpc} dk \left( R \left[ \alpha \left( \eta \right) ; k, \eta \right] - 1 \right)^2 } \,, 
\label{error-E}
\end{equation} 
where the lower integration limit $k \geq 0.1 \,\hmpc$ is set to a value at which the contribution to the source starts to be relevant, while the upper integration limit  $k \leq 0.4 \,\hmpc$ is about the maximal value to which our method provides a $\sim {\rm few} \; \%$ relative error. 
In our computation we will produce the nonlinear PS at the denominator of \re{ratio-R} by using the publicly available Coyote Emulator \cite{Lawrence:2009uk,Heitmann:2013bra},  whereas we will use our numerical simulations only for the computation of the source correlators at the last term at the numerator. Indeed, since the latter terms become relevant only at rather small scales (that is, for $k$ larger than $~ 0.1\, \hmpc$) we do not need large volume simulations to compute them reliably in a short time. On the other hand, these simulations are not optimal for the computation of the nonlinear PS at larger scales, where in our approach the dominant contribution is given by perturbative terms. The Coyote emulator, being originated by large volume and high accuracy N-body simulations, provides an useful tool to produce the nonlinear PS over the full range of scales we are interested in.  However, we checked that the fully non-linear PS from our simulations is in good agreement with the one from the Coyote Emulator.

\begin{figure}[ht!]
\centerline{
\includegraphics[width=0.55\textwidth,angle=0]{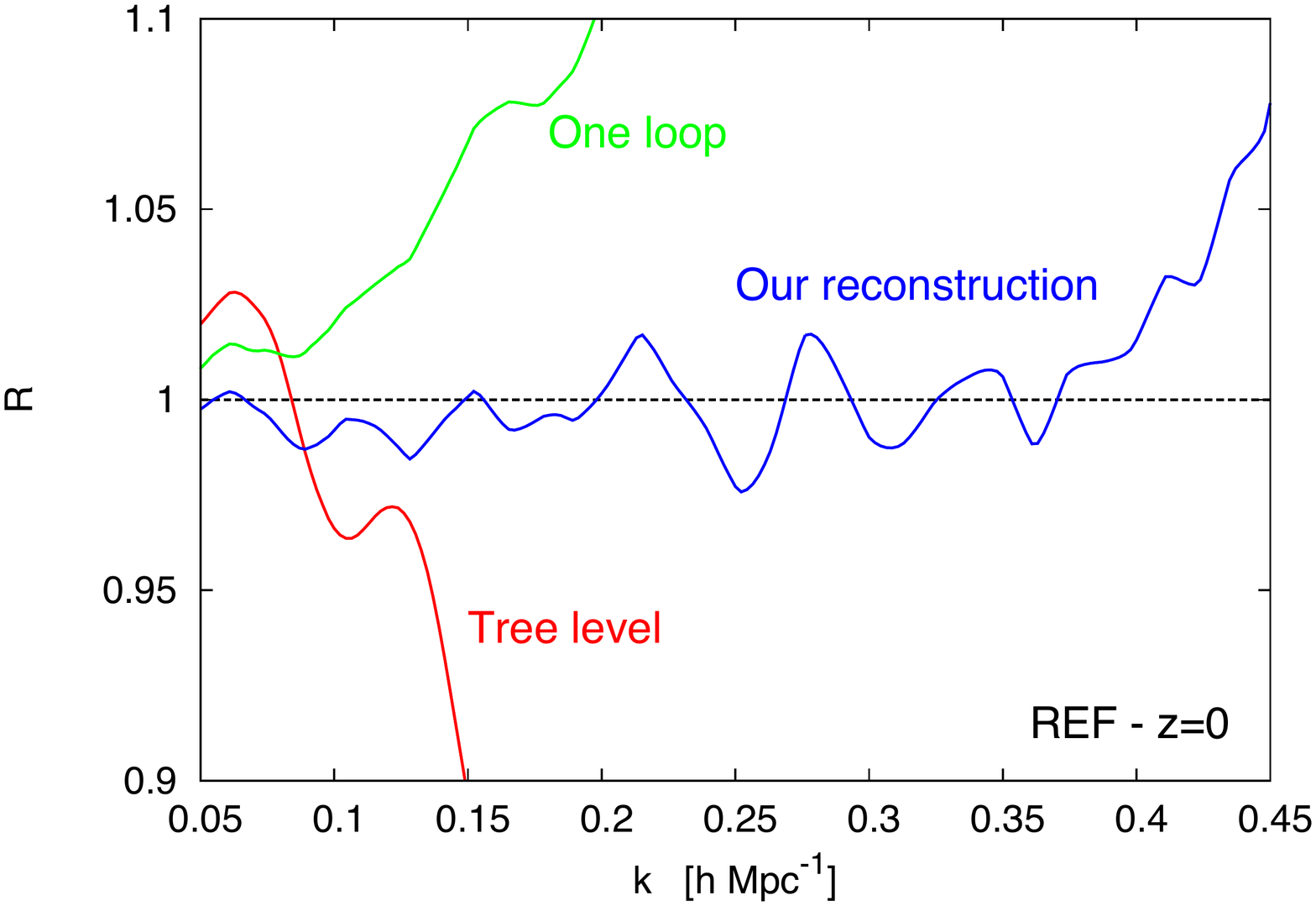}
\includegraphics[width=0.55\textwidth,angle=0]{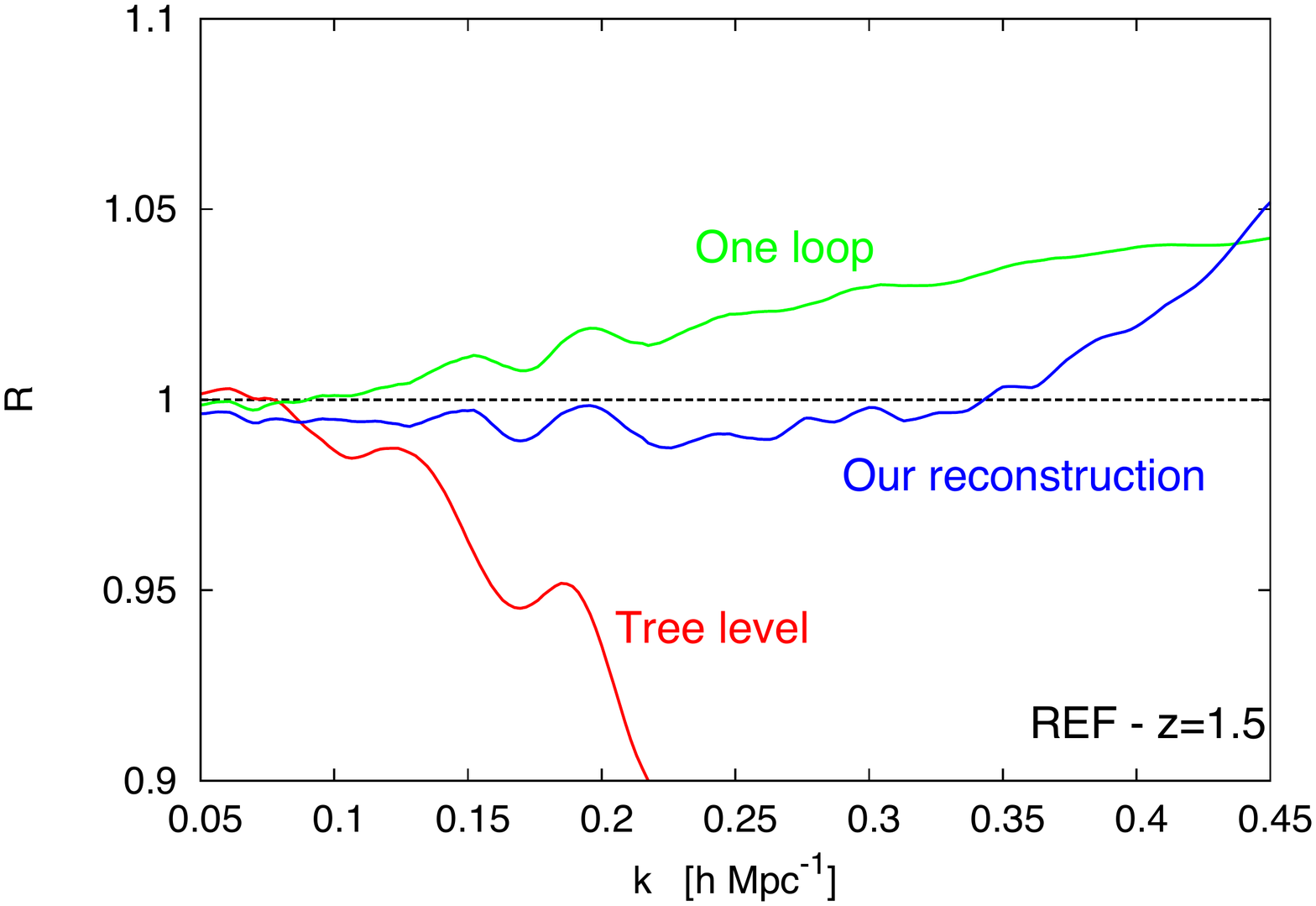}
}
\caption{
Ratio in eq. \re{ratio-R} between the reconstructed density PS and the full PS from the N-body simulations, 
at $z=0$ (left panel) and at $z=1.5$ (right panel).  The curved labelled ``Tree level'' 
contains only the linear tree level PS at the numerator; the curved labelled ``One loop'' contains the 1-loop PS of standard perturbation theory. The curve labelled ``Our reconstruction'' contains the full numerator, corresponding to the full expression for our reconstructed PS, eq. \re{P-reconstructed} with the approximation \re{P-reconstructed2}. This curve is shown for the choice of $\alpha$ that minimizes the relative error of eq. \re{error-E}. 
}
\label{fig:rec-alpha}
\end{figure}

In Fig. \ref{fig:rec-alpha} we show the ratio \re{ratio-R} as a function of comoving momentum $k$ for the REF cosmology (see Table~\ref{tab_sims}) at the redshifts $z=0$ (left panel) and $z=1.5$ (right panel). We note that the reconstruction with the full expression eq. \re{P-reconstructed} with the approximation \re{P-reconstructed2} works significantly better than 1-loop SPT, resulting in an average relative error at the $\sim 1\%$ level in the $ 0.1 \hmpc \leq k \leq  0.4 \hmpc$ interval.

\begin{table}
\begin{center}
\begin{tabular}{|l  | l | l | l | l | }
  \hline                       
{\bf REF  } $z$ &  $\alpha_{\rm rec.} $ & $ {\rm err.}_{\rm rec.}$  & 
 $ {\rm err.}_{\rm 1-loop,stand.}$ &  $\alpha_{\rm scaling} $    \\ \hline
$0$ &  $1.76_{-0.05}^{+0.06}$ & $1.0\%$ & $15\%$ & $1.81$ \\ \hline
$0.25$ &  $1.81 _{-0.08}^{+0.08} $ & $1.2\% $ & $12\%$ & $1.82$ \\ \hline
$0.5$ &  $1.88_{-0.11}^{+0.12}$ & $1.3\% $ & $8.5\%$ & $1.85$ \\ \hline
$1$ &  $2.00_{-0.14}^{+0.16}$ & $1.0\%$ & $4.7\%$  & $1.92$ \\ \hline
$1.5$ &  $2.08_{-0.16}^{+0.19}$ & $0.8\%$ & $2.4\%$  &  \\ \hline
\end{tabular}
\caption{$\alpha_{\rm rec.}$ is the value of $\alpha$ in \re{ratio-R} that minimizes the relative error \re{error-E} between the N-body and our reconstructed density PS. This error, denoted by ${\rm err.}_{\rm rec.}$, can be immediately compared with the error obtained from standard perturbation theory at 1-loop. The range of $\alpha$ shown in the second column is the interval of values for which the error is $< 2 \times {\rm err.}_{\rm rec.}$. The last column shows the value of $\alpha$ deduced from the time evolution of the $h_2 \vpb_1$ correlator at large scales, see eq. \re{alpha-scaling}. 
}
\label{tab:REF}
\end{center}
\end{table}

Table~\ref{tab:REF} shows that this accuracy is obtained at all the redshifts that we have studied. The second column in the table indicates the value of $\alpha$ that minimizes the relative error \re{error-E}. We denote this quantity as $\alpha_{\rm rec}$. We denote by ${\rm err.}_{\rm rec.}$ the value of the relative error at $\alpha=\alpha_{\rm rec.}$. We show this quantity in the third column. In the second column we also provide ``error bars'' on the determination of $\alpha$. They are defined from the interval in $\alpha$ for which the error remains below $2 \times {\rm err.}_{\rm rec.}$. The fourth column of the table shows the average error in the same  $ 0.1\, \hmpc \leq k \leq  0.4\, \hmpc$ interval obtained with the standard 1-loop computation.  The comparison between the third and the fourth column immediately quantifies the improvement of our reconstruction with respect to the standard 1-loop result. Finally, in the last column of the table, we show the value of $\alpha$ obtained from its scaling with $D$. Specifically, in analogy with eq. \re{alpha}
(and using the fact that $\vp \simeq \vp^{lin} \simeq {\rm const.}$ at large scales), we define
\begin{equation}
\alpha_{\rm scaling} \equiv {\rm ln } \frac{\langle h_2 \left( z \right) \, \vpb_1 \left( z \right) \rangle'}{\langle h_2 \left( 1.5 \right) \, \vpb_1 \left( 1.5 \right) \rangle'} \Bigg/  {\rm ln } \frac{ D \left( z \right) }{ D \left( 1.5 \right) } \;, 
\label{alpha-scaling}
\end{equation}
where, for definiteness, the reference redshift is taken at $z=1.5$ and where the correlators are evaluated at 
the linear scale $k = 0.00543 \, \hmpc$ (this is the 4th largest scale in our grid). We see from the table that $\alpha_{\rm rec.}$ and  $\alpha_{\rm scaling}$ are in very good agreement with each other and, even using the values of  $\alpha_{\rm scaling}$ in \re{ratio-R}, the error is below $2\%$.

\begin{figure}[ht!]
\centerline{
\includegraphics[width=0.7\textwidth,angle=0]{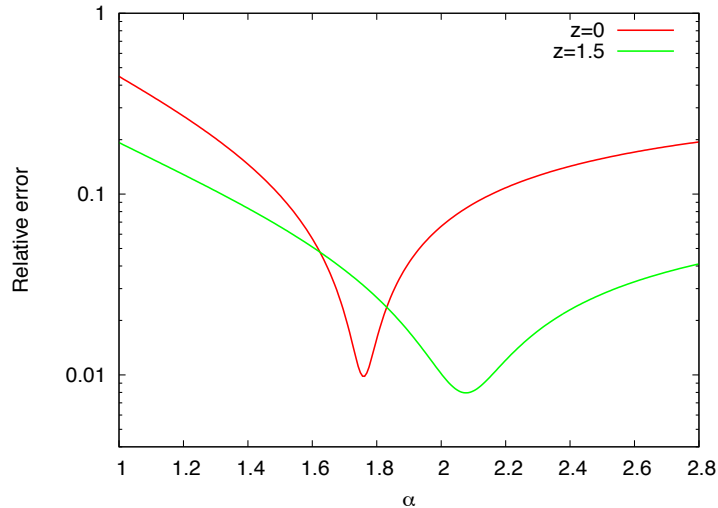}
}
\caption{
Relative error ${\cal E}$, given by eq. \re{error-E}, between  between the reconstructed density PS and the full PS from the N-body simulations, as a function of the parameter $\alpha$ and for two different redshift. 
}
\label{fig:erro}
\end{figure}

Finally, in Figure~ \ref{fig:erro} we show the relative error \re{error-E} as a function of $\alpha$ for two different values of the redshift. 

Let us briefly discuss these results. At the highest redshifts shown the nonlinear effects are less important, and  1-loop SPT is accurate at $\sim {\rm few} \%$ level in the range of $k$ that we are considering. The contribution from the source allows to improve the accuracy to $\la 1 \%$ level. At such redshifts, the coefficient $\alpha$ is consistent with $2$, which is the expected value in SPT, cf. the last two expressions in \ref{app-explicit}. As the redshift decreases, the accuracy of 1-loop SPT worsens to $> 10\%$, while the accuracy of our method remains at the $\sim 1-1.5 \%$. The value of $\alpha \left(z \right)$ decreases from the perturbative value, $\alpha=2$, signaling that the contribution from the UV becomes less important than the 1-loop SPT.\footnote{This may be understood as an effect of the decoupling of short scales consequent to their virialization \cite{Baumann:2010tm},  which would induce an effective UV cutoff, decreasing with $z$, in the loop integrals. A more thorough analysis is needed to verify this interpretation.} We also see that the range of $\alpha$ where the reconstruction works becomes narrower at low redshifts. This is an indication that the impact of the source becomes more relevant at these redshifts, so that a mistake in its characterization (namely, in its evolution encoded by $\alpha$) results in a greater penalty in the reconstruction. 

\section{Cosmology dependence}
\label{cosmodep}
In the previous section we have seen that the nonlinear PS from N-body simulations can be reproduced at the percent level up to $k\sim 0.4\, \hmpc$ by using the expression at the RHS of eq.~\re{P-reconstructed}. The first two terms in that expression are 1-loop quantities, which can be computed very rapidly for any cosmology. On the other hand, the third quantity, namely, $ \Delta \bar P_{11}^{h,{\rm N-body}}$, is obtained from an N-body simulation, which typically requires much longer computing times. 

A fully legitimate question is of course why the use of the RHS of eq.~\re{P-reconstructed} should be convenient with respect to directly using its LHS, that is, taking the nonlinear PS entirely from simulations. One first answer is that since the $\Delta \bar P_{11}^{h,{\rm N-body}}$ term mostly matters at small scales, it typically requires smaller simulation volumes, and less realizations, than a simulation aiming at reproducing the PS at all scales, including the linear or almost linear ones. In our approach, these scales are treated by PT, while the N-body simulations are employed only for the scales in which they are really needed. 

An even more attractive possibility would be to be able to parameterize the dependence of the  $\Delta \bar P_{11}^{h,{\rm N-body}}$ term on the cosmology and on time in a simple way, so that, once it is measured carefully from a given N-body simulation, it can be easily ``translated" to another cosmology and/or redshift. If this would be possible, then it would open the way to a method for the systematic scanning of different models in the nonlinear regime. In this and in the next section we show that this is indeed possible. 

First of all, we repeat the procedure described in the previous section to obtain the exponent in eq.~\re{alpha} from the fit of eq.~\re{P-reconstructed} to the output of the Coyote emulator, for all the other cosmologies in our suite, see Table~\ref{tab_sims}.
 The results are given in Table~\ref{tab:cosmoalpha}, from which we see that percent accuracies can be obtained also for these cosmologies.

\begin{table}
\begin{tabular}{|l|l|l|l||l|l|l||l|l|l|}  \hline                       
$z$ & ${\bf A_s^-:}$  $\alpha_{\rm rec.} $ & $ {\rm err.}$  & $\alpha_{\rm scal.} $ & 
${\bf A_s^+:}$  $\alpha_{\rm rec.} $ & $ {\rm err.}$  &  $\alpha_{\rm scal.} $ & 
${\bf n_s^-:}$  $\alpha_{\rm rec.} $ & $ {\rm err.}$  &  $\alpha_{\rm scal.} $     \\ \hline
$0$ & $1.83_{-0.06}^{+0.06}$ & $0.9\%$ &  $1.86$ & 
$1.61_{-0.05}^{+0.05}$ & $1.2\%$ &   $1.70$ & 
 $1.86_{-0.05}^{+0.05}$ & $0.8\%$   & $1.87$  \\ \hline 
$0.25$ & $1.88_{-0.08}^{+0.09}$ & $0.9\%$ &   $1.88$ &  
$1.66_{-0.07}^{+0.07}$ & $1.4\%$ &   $1.74$ & 
$1.91_{-0.06}^{+0.07}$ & $0.7\%$  &  $1.89 $     \\ \hline
$0.5$ & $1.96_{-0.10}^{+0.11}$ & $0.9\%$ &    $1.95$ & 
$1.74_{-0.09}^{+0.10}$ & $1.4\%$ &   $1.75$ & 
$1.98_{-0.09}^{+0.10}$ & $0.8\%$  &  $1.93 $     \\ \hline
$1$ & $2.04_{-0.12}^{+0.14}$ & $0.7\%$ &    $2.00$ & 
$1.85_{-0.11}^{+0.13}$ & $1.2\%$ &   $1.80$ & 
$2.07_{-0.11}^{+0.12}$ & $0.7\%$  &  $1.99 $     \\ \hline
$1.5$ & $2.09_{-0.14}^{+0.17}$ & $0.6\%$ &     & 
$1.95_{-0.13}^{+0.15}$ & $0.9\%$ &    & 
 $2.11_{-0.12}^{+0.14}$ & $0.5\%$  &         \\ \hline \hline 
$z$ & ${\bf n_s^+:}$  $\alpha_{\rm rec.} $ & $ {\rm err.}$  & $\alpha_{\rm scal.} $ & 
${\bf \Omega_m^-:}$  $\alpha_{\rm rec.} $ & $ {\rm err.}$  & $\alpha_{\rm scal.} $ & 
${\bf \Omega_m^+:}$  $\alpha_{\rm rec.} $ & $ {\rm err.}$   &  $\alpha_{\rm scal.} $     \\ \hline 
$0$ & $1.59_{-0.05}^{+0.06}$ & $1.3\%$ &   $1.70$ & 
$1.92_{-0.07}^{+0.07}$ & $1.0\%$ &   $1.90$ &  
 $1.60_{-0.07}^{+0.08}$ & $1.7\%$    &  $1.72 $     \\ \hline
$0.25$ & $1.65_{-0.08}^{+0.08}$ & $1.5\%$ &   $1.75$ & 
$2.01_{-0.13}^{+0.15}$ & $1.4\%$ &   $1.91$ & 
 $1.65_{-0.07}^{+0.07}$ & $1.3\%$  &  $1.74 $     \\ \hline
$0.5$ & $1.73_{-0.10}^{+0.11}$ & $1.5\%$ &   $1.73$ & 
$2.10_{-0.22}^{+0.28}$ & $2.0\%$ &   $1.94$ & 
 $1.71_{-0.07}^{+0.08}$ & $1.1\%$  &  $1.76 $     \\ \hline
$1$ & $1.85_{-0.13}^{+0.15}$ & $1.3\%$ &   $1.79$ & 
$2.23_{-0.31}^{+0.45}$ & $1.8\%$ &   $2.01$ & 
 $1.81_{-0.08}^{+0.08}$ & $0.7\%$   &  $1.79 $     \\ \hline
$1.5$ & $1.95_{-0.15}^{+0.18}$ & $1.0\%$ &   & 
$2.30_{-0.34}^{+0.51}$ & $1.3\%$ &    & 
 $1.88_{-0.09}^{+0.10}$ & $0.6\%$   &       \\ \hline
\end{tabular}
\caption{Parameter $\alpha$ obtained from the best fit in eq. \re{ratio-R} for the $A_s^\pm, \; n_s^\pm$, and $\Omega_m^\pm $ cosmologies listed in Table~\re{tab_sims}, and for several redshifts. 
}
\label{tab:cosmoalpha}
\end{table}

\begin{figure}[ht!]
\centerline{
\includegraphics[width=0.55\textwidth,angle=0]{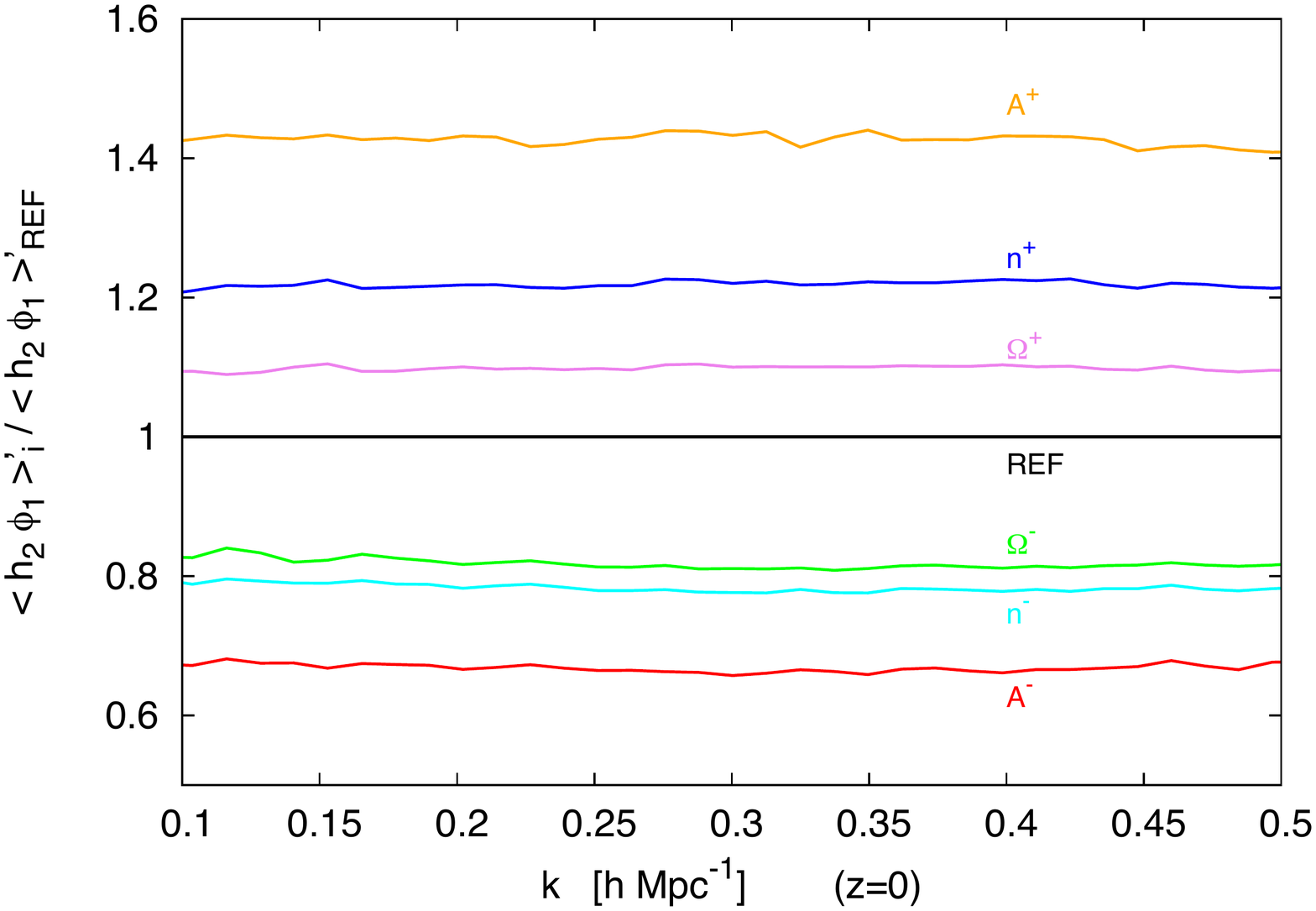}
\includegraphics[width=0.55\textwidth,angle=0]{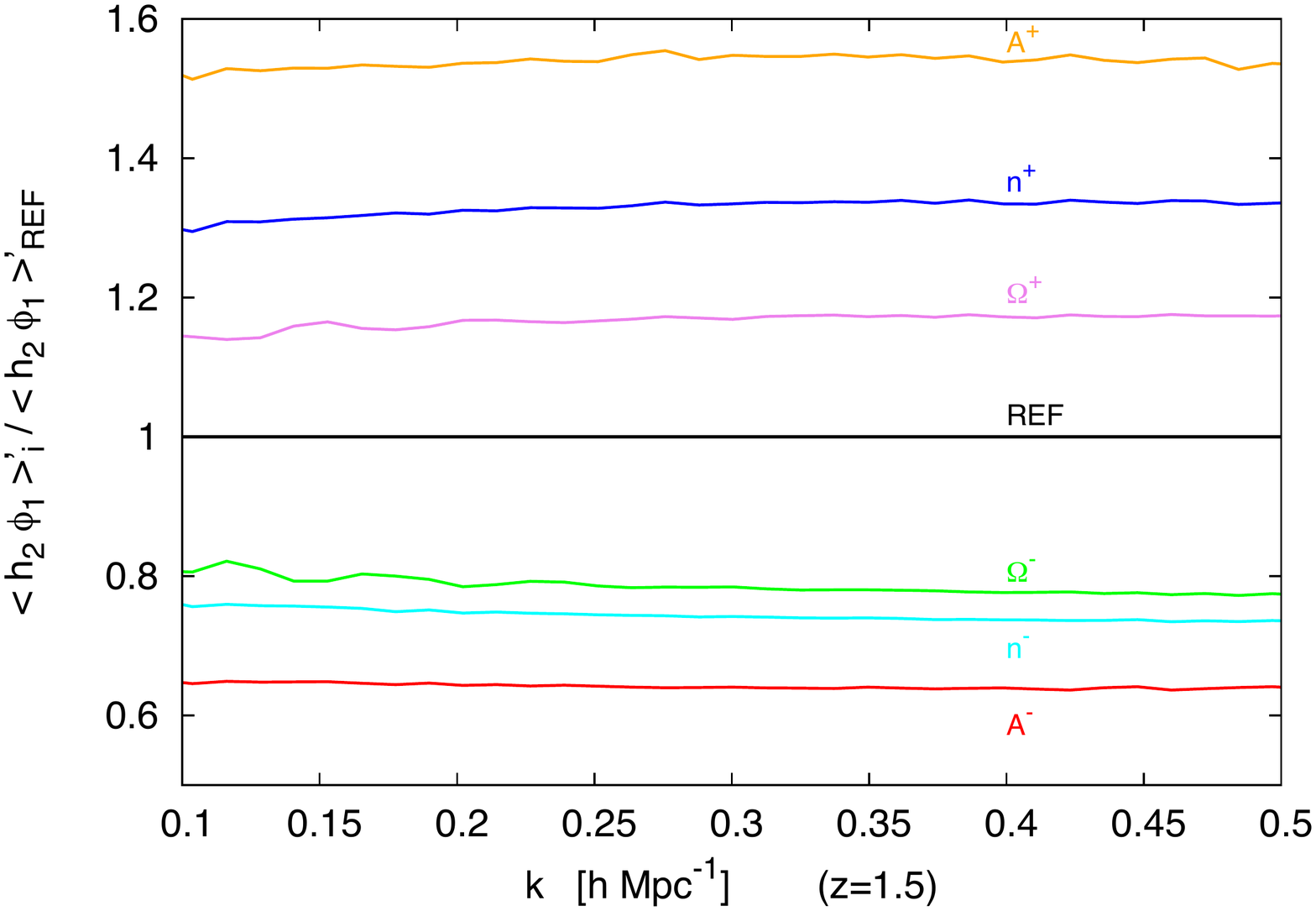}
}
\caption{
Ratios $\left\langle  h_2  \left( \bk, \eta \right)    \vpb_{1}(\bk',\eta) \right\rangle'_{i}/\left\langle  h_2  \left( \bk, \eta \right)    \vpb_{1}(\bk',\eta) \right\rangle'_{REF}$ for the cosmologies in Table~\ref{tab_sims} and for two different redshifts. The label on each line indicates the cosmology at the numerator. 
}
\label{fig:simple_ratios}
\end{figure}

Next, we discuss the truly non-perturbative contribution to eq.~\re{P-reconstructed}, namely, of $\Delta \bar P_{11}^{h,{\rm N-body}}$.
From eq.~\re{P-reconstructed2} we see that its scale dependence is entirely encoded in the correlator $ \left\langle  h_2  \left( \bk, \eta \right)    \vpb_{1}(\bk',\eta) \right\rangle' $.  In Figure~\ref{fig:simple_ratios} we plot the ratios between these correlators computed in the cosmologies of Table~\ref{tab_sims} and the one computed in the reference cosmology, for two  different redshifts. As we see, all the ratios are approximately constant in the whole range of relevant scales. The curves for the cosmologies $\Omega_m^+$ and $\Omega_m^-$ exhibit some wiggles in the BAO range of scales, which are expected from the fact that, already at the linear level, the BAO peaks are shifted in these cosmologies with respect to the REF one. On the other hand, the other four cosmologies in  Table~\ref{tab_sims} have the same growth function and $H(z)$ as the REF cosmology, and therefore the same  peak positions. In principle, the residual wiggles in the $\Omega_m^+$ and $\Omega_m^-$ curves can be eliminated, or reduced, by using the linear information about the peak shifts, however, in this paper we will not do so since the residual scale dependence does not worsen the accuracy of the reconstruction above the percent level (see below).

\begin{figure}[ht!]
\centerline{
\includegraphics[width=0.55\textwidth,angle=0]{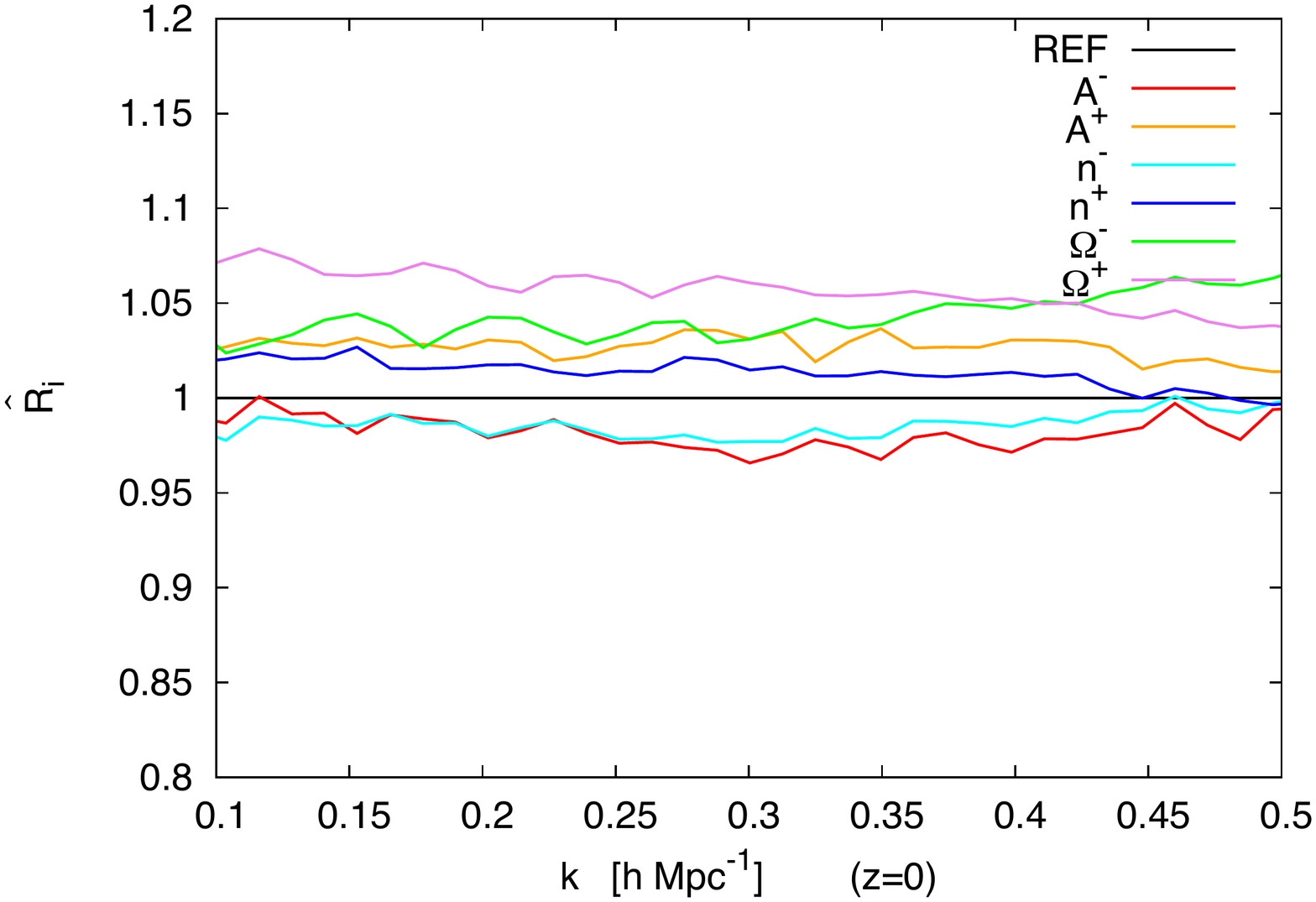}
\includegraphics[width=0.55\textwidth,angle=0]{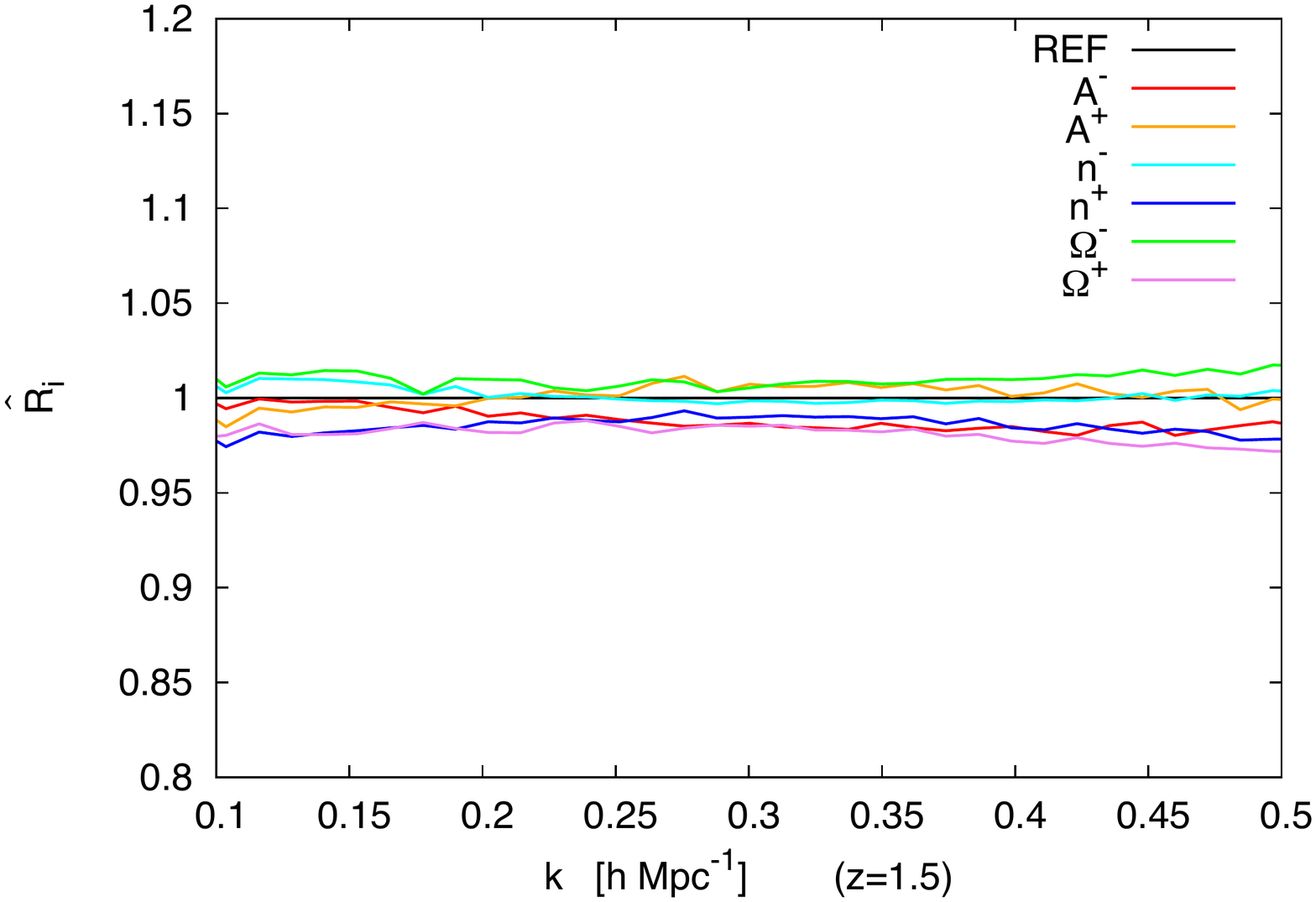}
}
\caption{
Ratios ${\hat R}_i$ in eq. \re{R-hat} for two different redshifts. Notice the range in the $y-$axis is significantly smaller than the one in Figure~ \ref{fig:simple_ratios} and that the ratios ${\hat R}_i$ differ from one by at most a few percent. 
}
\label{fig:Ri_ratios}
\end{figure}

Due to the approximate scale invariance of the $\frac{\langle \vpb_1 h_2 \rangle'_i}{\langle \vpb_1 h_2 \rangle'_{\rm REF}}$, we can in principle employ the $\langle \vpb_1 h_2\rangle'_{\rm REF}$ correlator for the reconstruction of the PS of any other cosmology up to an overall (i.e. $k-$independent) rescaling factor, without the need of computing the source from an N-body simulation for that cosmology. 
 We now show that this rescaling factor, and actually an even better $k-$independence, can be obtained if we also employ information encoded both in SPT and in the time dependence of eq.~\re{alpha}.  More specifically, we will consider, for each cosmology, the ratio
\beq
{\cal R}_i(k,\eta)\equiv \frac{ \langle h_2 \left( \bk, \eta \right)   \vpb_{1}(\bk',\eta) \rangle_i' }{ \Delta \bar P_{ss,11,i}^{h,{\rm 1-loop}} (k,\eta)}.
\eeq
The time dependence of the numerator is given by eq.~\re{alpha}, while that of the denominator is simply $\propto D_i^2$, therefore 
\beq
{\cal R}_i(k,\eta) \sim {\cal R}_i(k,\eta_*) \left(\frac{D_i(\eta)}{D_i(\eta_*)}\right)^{\alpha_i(\eta)-2}\,,
\eeq
where $\eta_*$ corresponds to a redshift at which $\alpha_i$ has converged to its SPT value, $\alpha_i (\eta_*)\simeq \alpha_{SPT}= 2$.  
If this is the case, ${\cal R}_i \left( k , \eta \right) = {\cal R}_i \left( k , \eta_* \right) $ for all $\eta < \eta_*$. At very early time, we expect that the nonperturbative effects should be negligible, and therefore ${\cal R}_i \left( k , \eta \right) $ should be nearly cosmology- and momentum-  independent. Therefore, we expect the ratio
\begin{equation}
\widehat {\cal R}_i \left( k , \eta \right) \equiv 
\frac{{\cal R}_i \left( k, \eta \right)}{{\cal R}_{\rm REF} \left( k, \eta \right)} \, 
\frac{ \left( \frac{D_{\rm REF} ( \eta)}{D_{\rm REF} (\eta_*)} \right)^{\alpha_{\rm REF} \left( \eta \right) - 2} }{\left( \frac{D_i ( \eta)}{D_i ( \eta_* )} \right)^{\alpha_i \left( \eta \right) - 2}}  \sim
\frac{{\cal R}_i \left( k, \eta_* \right)}{{\cal R}_{\rm REF} \left( k, \eta_* \right)},
\label{R-hat}
\end{equation}
to be approximately equal to one.
In Figure~~\ref{fig:Ri_ratios} we plot the ratios $\widehat{\cal R}_i$ for the same cosmologies and redshifts considered in Figure~ \ref{fig:simple_ratios}. We note that the rescaled ratios are much closer to one than those shown  in Figure~ \ref{fig:simple_ratios}.

\section{PS reconstruction for a new cosmology without running a new N-body}
\label{sec:noNbody}
In this section we will use the results of the previous one to put our method at work, namely to use it to predict a nonlinear PS for a given cosmology without actually running a N-body simulation for that cosmology. Using eq.~\re{R-hat} to relate the $\langle h \vpb \rangle$ correlators of different cosmologies, and \re{P-reconstructed2}, we obtain the predicted PS for the $i$-cosmology given the nonlinear correlator measured for the REF cosmology, $\langle h \vpb \rangle_{REF}$ as
\beqra
&&\!\!\!\!\! \!\!\!\!\!\!\!  \!\!\!\!\!\!\! \!\!\!\!\!\!\! \Delta \bar P_{11,i}^{h,{\rm N-body}} (k,\eta) \simeq  - \frac{4}{\alpha_i \left( \eta \right) \left[ 5 + 2 \alpha_i \left( \eta \right) \right] }  \left\langle 
 h_2  \left( \bk, \eta \right)    \vpb_{1}(\bk',\eta) \right\rangle_i'  \,,\nonumber\\
 &&\!\!\!\!\!\!\! \!\!\!\!\!\!\! \simeq  - \frac{4}{\alpha_i \left( \eta \right) \left[ 5 + 2 \alpha_i \left( \eta \right) \right] }  \,\frac{\left(\frac{D_i(\eta)}{D_i(\eta_*)}\right)^{\alpha_i(\eta)-2}}{\left(\frac{D_{REF}(\eta)}{D_{REF}(\eta_*)}\right)^{\alpha_{REF}(\eta)-2}}\,{\cal R}_{REF}(k,\,\eta)\,\Delta \bar P_{ss,11,i}^{h,{\rm 1-loop}} (k,\eta) \,,\nonumber\\
\eeqra
which is then inserted in \re{P-reconstructed} to obtain the nonlinear PS for the $i$-cosmology: 
\beqra
& & \!\!\!\!\!\!\!\!  \!\!\!\!\!\!\!\!  
 \bar P_{11,i} (k,\eta)  \simeq  \bar P_{ss,11,i} ^{\rm 1-loop} 
\nonumber\\ 
&  &\!\!\!\!\!\!\! \!\!\!\!\!\!\!  - \left\{ 1+ {\cal C}_i \left( \eta \right) \, \left( \frac{ D_{\rm REF} \left( \eta_* \right)}{D_{\rm REF} \left( \eta \right) } \right)^{\alpha_{\rm REF} \left( \eta \right) - 2 }  \frac{ \langle h_2 \left( \bk, \eta \right)   \vpb_{1}(\bk',\eta) \rangle_{\rm REF}' }{ \Delta \bar P_{ss,11,{\rm REF}}^{h,{\rm 1-loop}} (k,\eta)}  \right\}  \Delta \bar P_{ss,11,i}^{h,{\rm 1-loop}} (k,\eta) \,,
  \nonumber\\ 
\label{P-reconstructed3} 
\eeqra
where we have defined 
\beqra
 {\cal C}_i \left( \eta \right)  \equiv  \frac{4}{\alpha_i \left( \eta \right) \left[ 5 + 2 \alpha_i \left( \eta \right) \right] }  \, \left(\frac{D_i(\eta)}{D_i(\eta_*)}\right)^{\alpha_i(\eta)-2} \,. 
\label{calC}
\eeqra
The RHS of eq. \re{P-reconstructed3} constitutes our reconstruction for the density PS of any given cosmology (recall that the index $i$ labels any arbitrary cosmology).  Apart from the coefficient ${\cal C}_i \left( \eta \right)$, this RHS can be evaluated without the need to obtain the source for that cosmology from an N-body simulation.

When trying to see whether a set of density data is compatible with a given cosmology, we can insert \re{calC} into  \re{P-reconstructed3}, and leave $\alpha_i \left( \eta \right)$ as a free coefficient, which should be  fitted together with the data. The error interval for $\alpha$ in Table \re{tab:CGPT} (see below) shows that $\alpha$ does not need to be determined with a percent accuracy for the final PS to be percent accurate, but that in most cases even a $10\%-20\%$ accuracy is enough. Alternatively, one can evaluate and table the coefficients ${\cal C}_i \left( \eta \right)$ for a grid of cosmologies, and then interpolate.

We now apply this reconstruction procedure to the last $6$ cosmologies of Table~\ref{tab_sims}, pretending that we have not evaluated the sources $\langle h_2 \vpb_1 \rangle_i$ for them.  In this way we can infer the value of $\alpha_i \left( \eta \right)$ that minimizes the difference between the LHS and RHS of 
\re{P-reconstructed3}. We define this quantity as   $\alpha_{\rm CGPT} \left( \eta \right)$, and we define the error in this procedure, ${\rm err.}_{\rm CGPT}$, and the range $\pm  \Delta \alpha_{\rm CGPT} \left( \eta \right)$, precisely as we did for the analogous quantities in Table~\ref{tab:cosmoalpha}. We show these results in Table~4.

\begin{table}
 \label{tab:cosmoCGPT}
\centering
\begin{tabular}{|l|l|l||l|l||l|l|}  \hline                       
$z$ & ${\bf A_s^-:}$  $\alpha_{\rm CGPT} $ & $ {\rm err.}$  & 
${\bf A_s^+:}$  $\alpha_{\rm CGPT} $ & $ {\rm err.}$  & 
${\bf n_s^-:}$  $\alpha_{\rm CGPT} $ & $ {\rm err.}$    \\ \hline
$0$ & $2.18_{-0.50}^{+0.59}$ & $0.9\%$ & 
$1.47_{-0.14}^{+0.22}$ & $1.2\%$ &  
 $2.18_{-0.49}^{+0.56}$ & $0.8\%$  \\ \hline 
$0.25$ & $2.00_{-0.26}^{+0.63}$ & $0.9\%$ &  
$1.56_{-0.15}^{+0.20}$ & $1.4\%$ & 
$1.99_{-0.22}^{+0.41}$ & $0.8\%$    \\ \hline
$0.5$ & $2.02_{-0.23}^{+0.32}$ & $0.9\%$ & 
$1.67_{-0.17}^{+0.22}$ & $1.5\%$ & 
$2.01_{-0.20}^{+0.28}$ & $0.9\%$      \\ \hline
$1$ & $2.07_{-0.16}^{+0.20}$ & $0.7\%$ & 
$1.83_{-0.15}^{+0.18}$ & $1.2\%$ & 
$2.07_{-0.15}^{+0.17}$ & $0.7\%$   \\ \hline
$1.5$ & $2.11_{-0.14}^{+0.16}$ & $0.6\%$ &   
$1.94_{-0.13}^{+0.15}$ & $0.9\%$ &  
 $2.12_{-0.12}^{+0.13}$ & $0.5\%$         \\ \hline \hline 
$z$ & ${\bf n_s^+:}$  $\alpha_{\rm CGPT} $ & $ {\rm err.}$  & 
${\bf \Omega_m^-:}$  $\alpha_{\rm CGPT} $ & $ {\rm err.}$  & 
${\bf \Omega_m^+:}$  $\alpha_{\rm CGPT} $ & $ {\rm err.}$     \\ \hline 
$0$ & $1.53_{-0.17}^{+0.29}$ & $1.3\%$ & 
$2.25_{-0.58}^{+0.69}$ & $1.1\%$ & 
 $1.57_{-0.23}^{+1.59}$ & $1.6\%$     \\ \hline
$0.25$ & $1.62_{-0.18}^{+0.26}$ & $1.5\%$ &  
$2.3_{-0.5}^{+2.1}$ & $1.5\%$ &  
 $1.60_{-0.15}^{+0.21}$ & $ 1.2\%$     \\ \hline
$0.5$ & $1.71_{-0.19}^{+0.26}$ & $1.6\%$ & 
$2.3_{-0.5}^{+4.8}$ & $2.0\%$ &  
 $1.68_{-0.13}^{+0.16}$ & $1.0\%$    \\ \hline
$1$ & $1.86_{-0.16}^{+0.20}$ & $1.2\%$ & 
$2.27_{-0.43}^{+0.74}$ & $1.8\%$ & 
 $1.81_{-0.10}^{+0.11}$ & $0.7\%$     \\ \hline
$1.5$ & $1.96_{-0.15}^{+0.18}$ & $0.9\%$ &  
$2.29_{-0.34}^{+0.51}$ & $1.3\%$ &   
 $1.90_{-0.08}^{+0.09}$ & $0.5\%$        \\ \hline
\end{tabular}
\caption{Parameter $\alpha$ obtained from the Coarse Grained Perturbation Theory  reconstruction of the cosmologies $A_s^{\pm}, n_s^{\pm}, \Omega_m^{\pm}$. eq.  \re{P-reconstructed3} is used, which only requires the source for the REF cosmology from N-body (rather than the sources of all the cosmologies). 
 }  
\label{tab:CGPT} 
\end{table}

The comparison between the ``rec.'' quantities of Table~\ref{tab:cosmoalpha} and the ``CGPT'' ones of Table~\ref{tab:CGPT}   quantify how the procedure discussed in this section allows to reconstruct the sources in the $i-$cosmology without running the corresponding N-body simulation. The interval for $\alpha_{\rm CGPT} $ shown in the table  gives a measure of the mistake that can be tolerated  in the determination of $\alpha_{\rm CGPT} $ from fitting the data. We recall that this interval  indicates the values of $\alpha$ that, once inserted in \re{calC} and then in \re{P-reconstructed3} result in an error that is at most twice the one obtained for the best fit value $\alpha_{\rm CGPT} $.

The big ``error bars''  in Tab.~\ref{tab:CGPT}  indicate that eq.~\re{P-reconstructed3}  is only mildly dependent on the parameter $\alpha_i(\eta)$. As we already mentioned, this implies that the parameter $\alpha_i$ does not need to be determined with percent accuracy to obtain a percent accurate PS. For all cosmologies and for all redshifts the values in Table \ref{tab:CGPT} (for which, we recall,  $\langle h_2 \vpb_1 \rangle_i$ has not been computed)  are compatible with those obtained for the REF cosmology in Table~\ref{tab:REF} (for which  $\langle h_2 \vpb_1 \rangle_i$ has been computed). This indicates that the bulk of the cosmology dependence is captured by the perturbative term $\Delta \bar P_{ss,11,i}^{h,{\rm 1-loop}} (k,\eta)$ in eq.~\re{P-reconstructed3}, and in particular it confirms that percent accuracy can be obtained without the need of running $N$-body simulations for each cosmology.

\section{Time dependence of the $\alpha$ parameter.}
\label{sec:time}
We now discuss the time-dependence of the $\alpha_{i}(\eta)$ for the different cosmologies, as reported in Table~\ref{tab:REF} for the REF cosmology and in Table~\ref{tab:cosmoalpha} for the other ones. 

\begin{figure}[ht!]
\centerline{
\includegraphics[width=0.4\textwidth,angle=0]{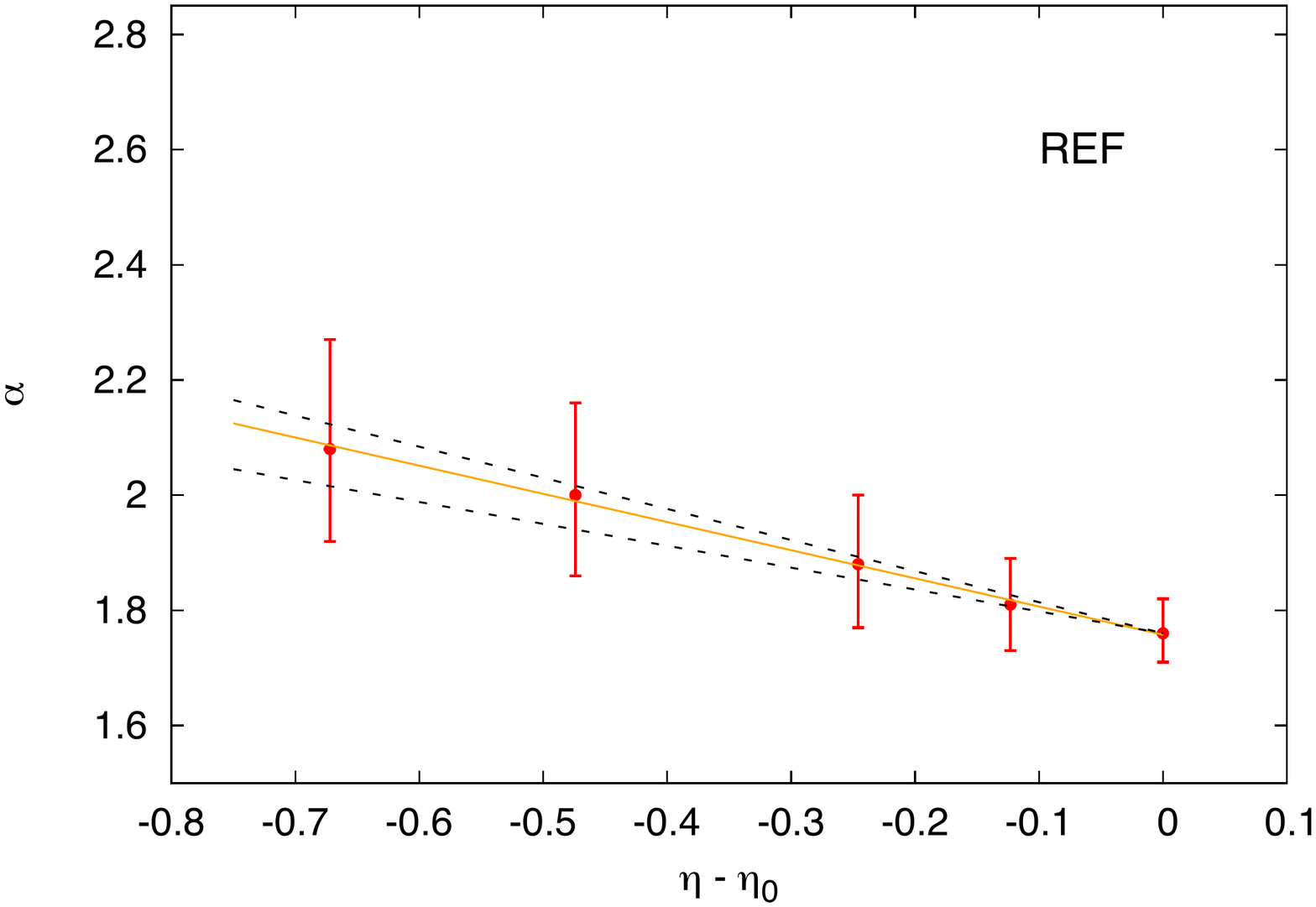}}
\centerline{\includegraphics[width=0.4\textwidth,angle=0]{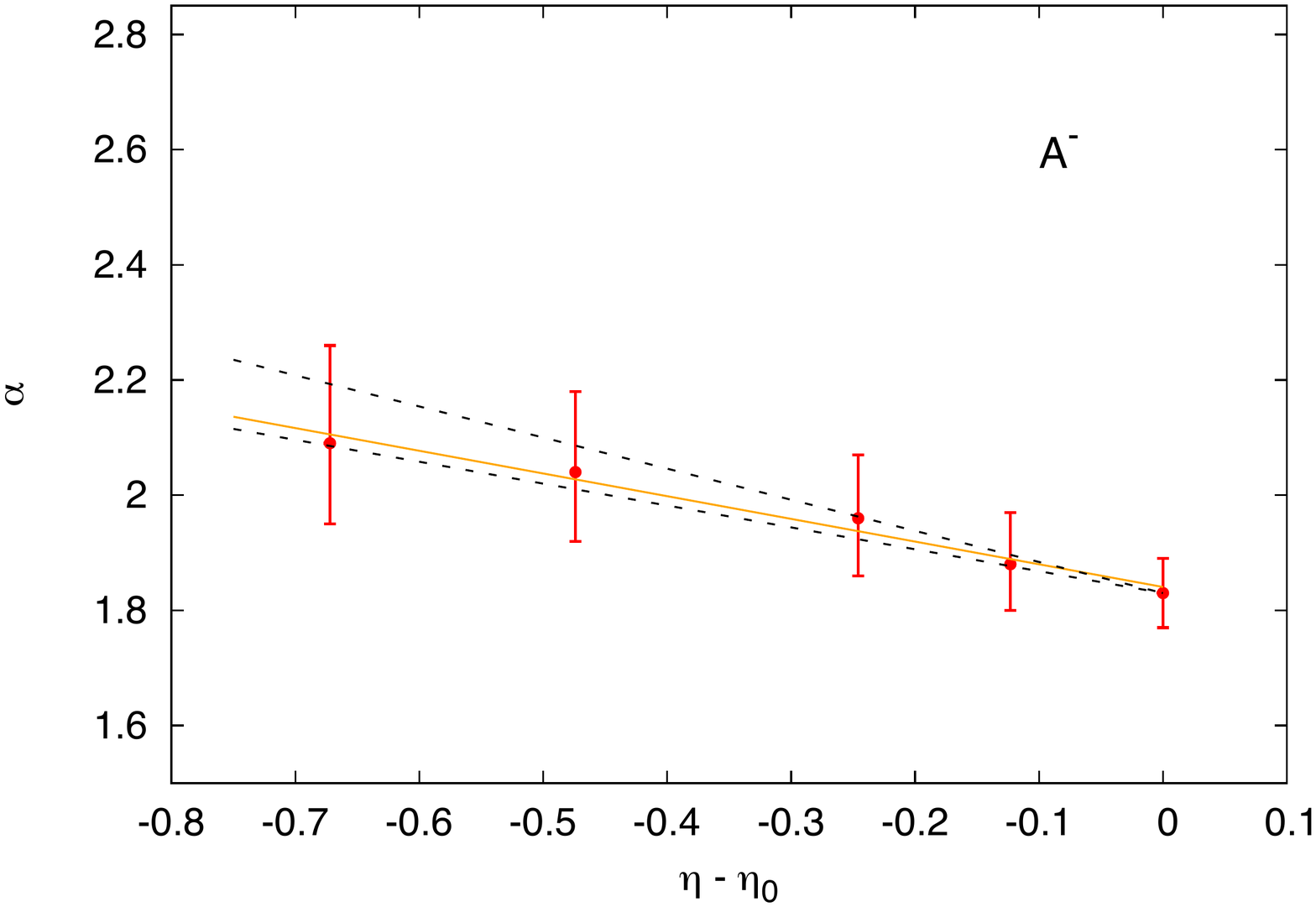} \includegraphics[width=0.4\textwidth,angle=0]{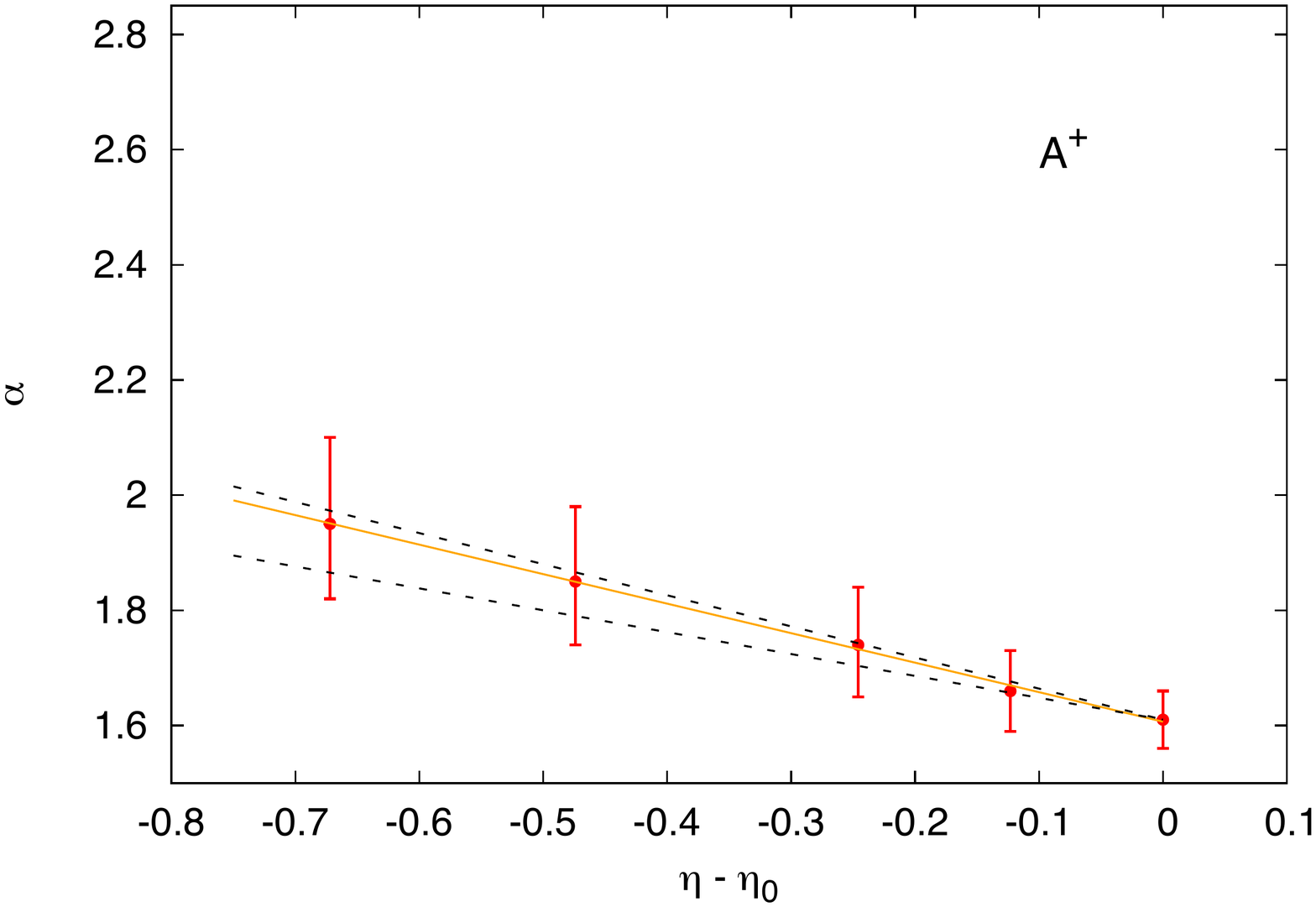}}
\centerline{\includegraphics[width=0.4\textwidth,angle=0]{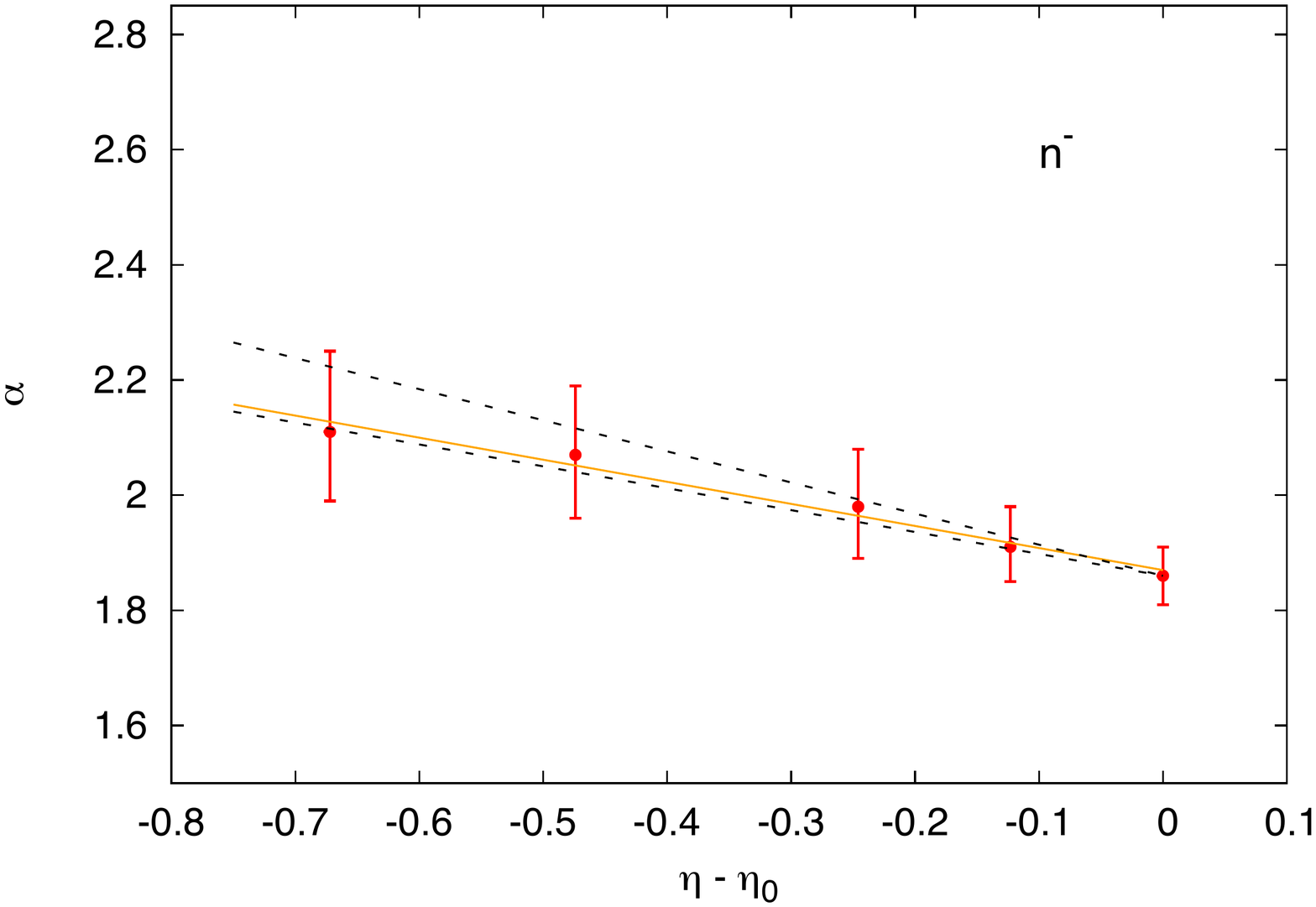} \includegraphics[width=0.4\textwidth,angle=0]{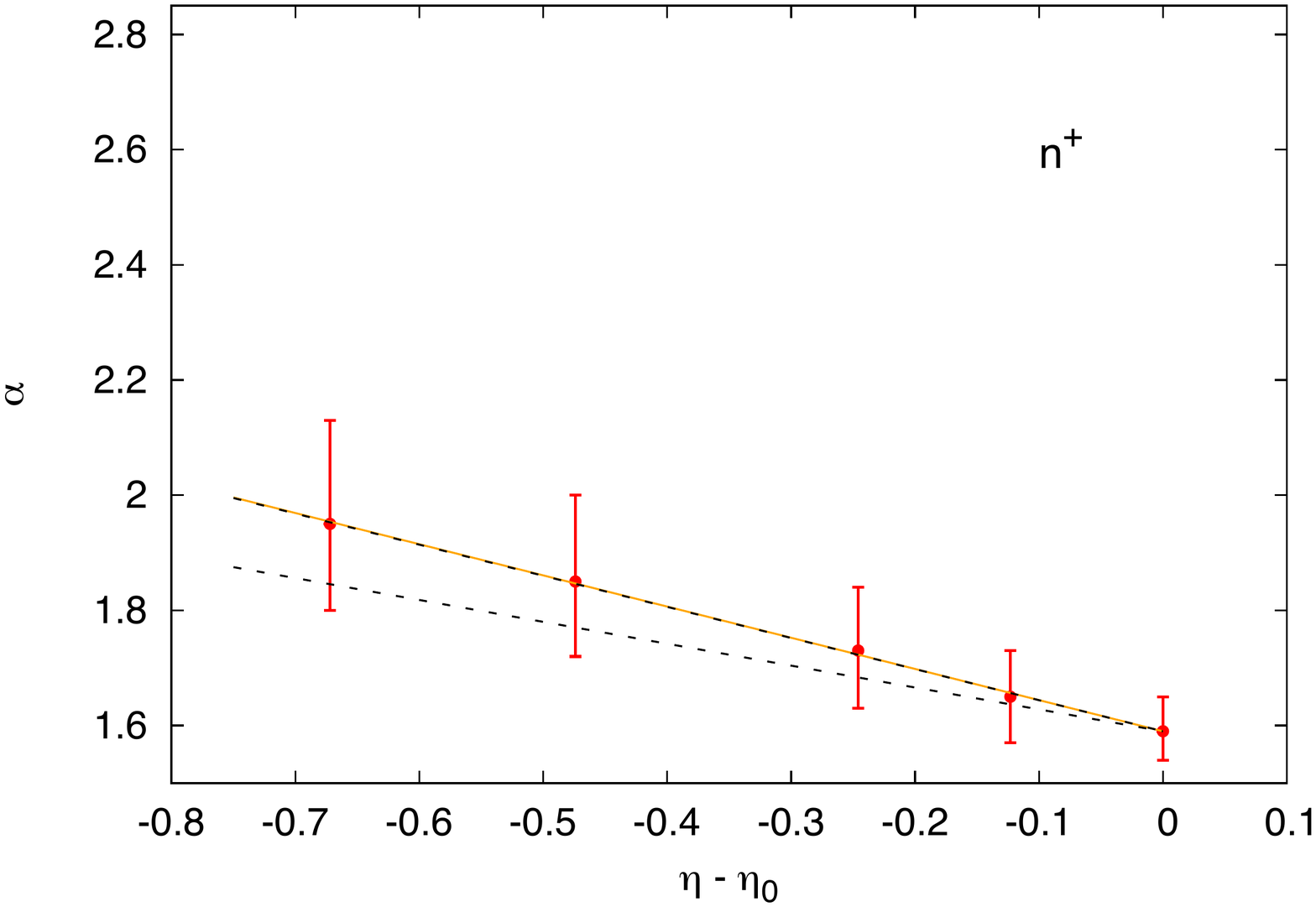}}
\centerline{\includegraphics[width=0.4\textwidth,angle=0]{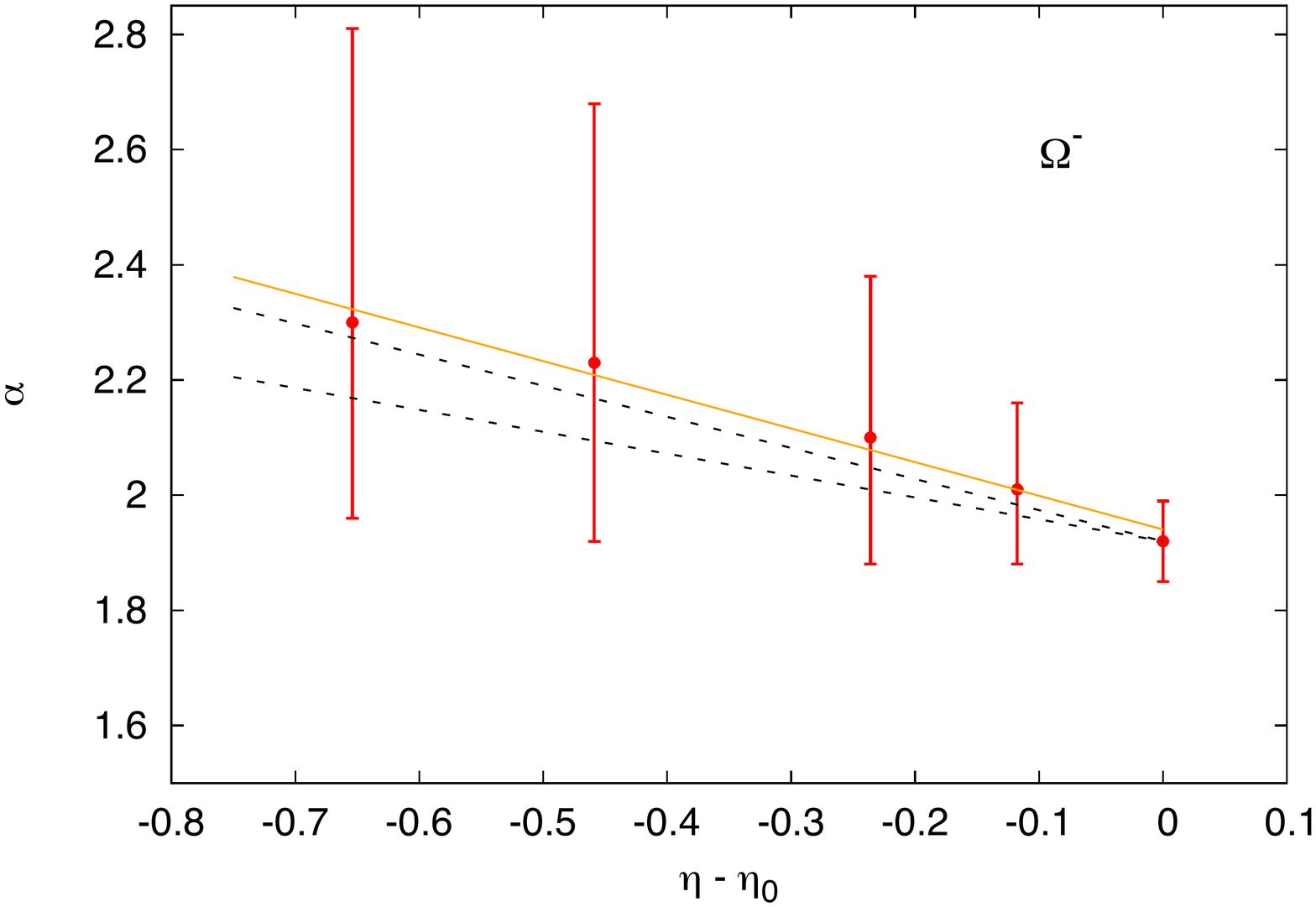} \includegraphics[width=0.4\textwidth,angle=0]{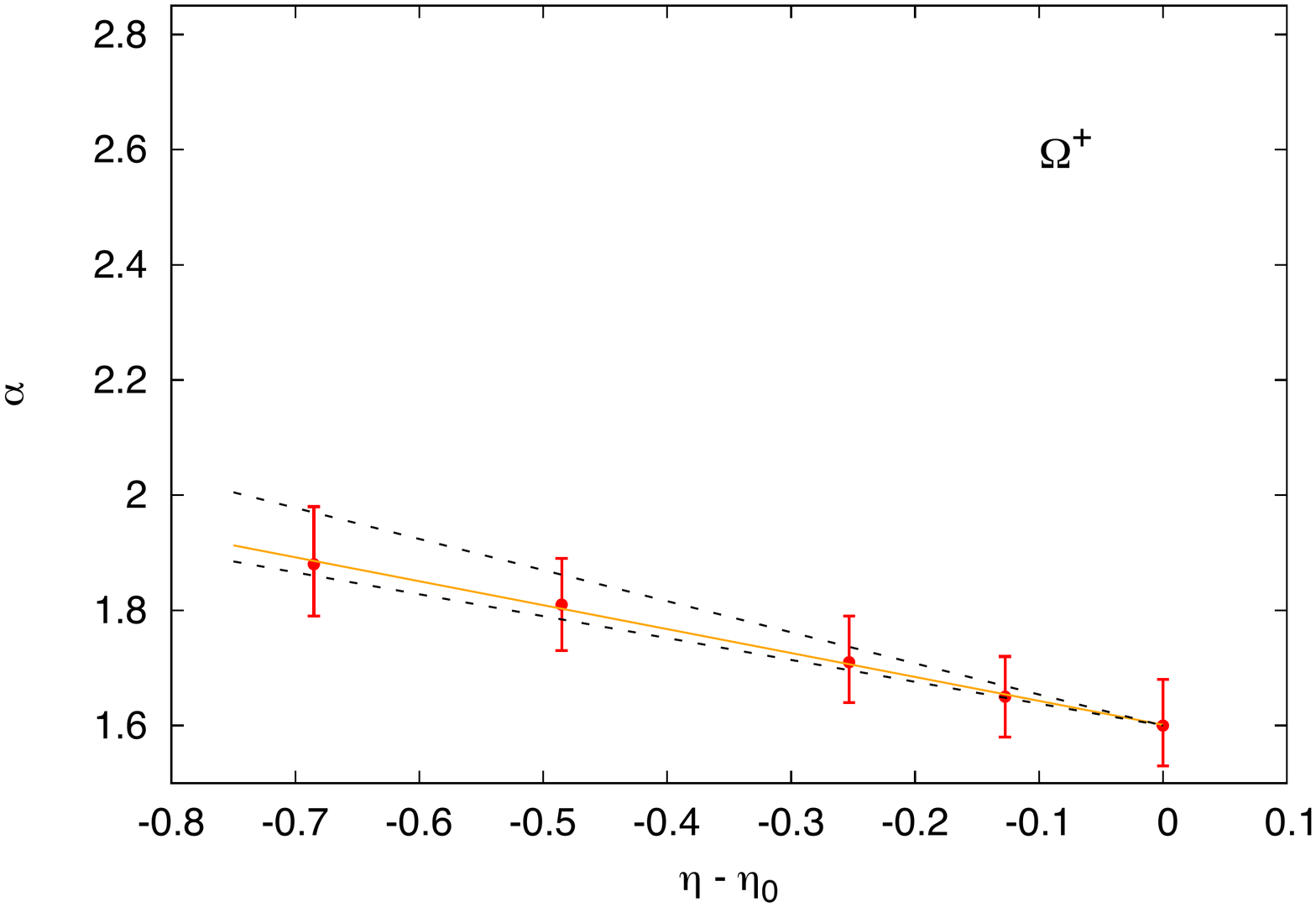}}
\caption{The time-dependent $\alpha_i$'s for the different cosmologies of our simulation suite. For each cosmology, the solid orange line indicates the best linear fit to the $\eta$ dependence, while the black dashed lines are obtained  from the largest ($\beta=0.58$)  and the smallest ($\beta=0.38$) slope in  Table~\ref{tab:betaslopes}. .
}
\label{fig:alphatime}
\end{figure}

In Figure~\ref{fig:alphatime} we plot the time-dependent values of Tables~\ref{tab:REF} and \ref{tab:cosmoalpha} (red points with error bars) and we notice that the central values are very well fit by a linear dependence in $\eta$ (red lines). The best fit slopes for each cosmology, defined as
\beq
\beta_i\equiv -\frac{d \alpha_{{\rm rec.},i}}{d \eta}\,,
\label{betai}
\eeq
where  $\alpha_{{\rm rec.},i}$ is the best linear fit to the $i$-th cosmology,  are given in Table~\ref{tab:betaslopes}. To gauge the sensitivity of our reconstruction procedure on the time-dependence of the $\alpha_i$'s we also plot, for each cosmology, the two dashed lines, corresponding to
\beq
\alpha_i^{min}(\eta) = \alpha_i(\eta_0) -(\eta - \eta_0)\,\beta^{min},\quad  \alpha_i^{max}(\eta) = \alpha_i(\eta_0) -(\eta - \eta_0)\,\beta^{max}\,,
\eeq
where $\beta^{min}=0.38$ ($\beta^{max}=0.58)$ is the smallest (largest) $\beta$ in   Table~\ref{tab:betaslopes}. As one can notice, taking any value of the slope between these two extrema gives values of $\alpha_i$ inside the error bars at any redshift.  This result is very encouraging, as it shows that once $\alpha_i$ is fixed at $z=0$ in a given cosmology, either by direct fitting or by the procedure described in the previous section, its values at higher redshifts can be reproduced accurately enough by assuming a linear dependence with the slope provided by, e.g. the REF cosmology. This provides further confirmation of the robustness of our Ansatz for the time dependence of the correlators, eq.~\re{alpha} and, more practically, it  shows that we do not need extra parameters to characterize the cosmological dependence of the time evolution.

\begin{table}
\centering
\begin{tabular}{| c | c | c | c | c | c | c | c | }
\hline
Cosmology & REF & $A_s^-$ & $A_s^+$ & $n_s^-$ &  $n_s^-$ & $\Omega_m^-$ & $\Omega_m^+$   \\
\hline \hline
 $\beta_i$ &  0.49&  0.39  & 0.51 & 0.38  & 0.54  & 0.58 & 0.42\\
\hline
\end{tabular}
\caption{Slopes of the best linear fit to $\alpha(\eta)$ for the different cosmologies, defined in eq.~\re{betai}.}
\label{tab:betaslopes}
\end{table}

\section{Comparison with EFT approach}
\label{EFToLSS}

In this section we discuss the relationship between our approach and the Effective Field Theory of Large Scale Structure (EFToLSS) approach of   \cite{Baumann:2010tm,Carrasco:2012cv,Carrasco:2013mua}.  Generally speaking, while our approach takes the full information on the small scale physics from N-body simulations, the EFToLSS uses the symmetries of the problem (and the assumption of gaussian initial coniditons) to express the effective stress tensor (which is closely related to our source term) as an expansion in long wavelength fields and their derivatives, up to so-called ``stochastic terms" (which are usually negelcted using the argument that their contribution to the PS scales only as $k^4$). Since this expansion is local in space, its coefficients are assumed to be scale independent, but are generally time-dependent and even non-local in time.

To put the comparison on a more concrete level, we start from eq. (39), 
\beqra
&&\!\!\! \!\!\! \!\!\! \langle \vpb_a(\bk,\eta) \vpb_b(\bk',\eta') \rangle = \langle \vpb_{ss,a}(\bk,\eta) \vpb_{ss,b}(\bk',\eta') \rangle \nonumber\\
 &&\;\;\;\;\;  + \langle \vpb_{\sigma,a}^{(1)}(\bk,\eta) \vpb_{ss,b}(\bk',\eta') \rangle +
 \langle \vpb_{ss,a}(\bk,\eta) \vpb_{\sigma,b}^{(1)}(\bk',\eta') \rangle +O(\langle \sigma^2\rangle)\,,
\eeqra
in which, we recall, the first term at the RHS is the fully nonlinear PS in the single stream approximation, while the other two give the effect of small scale velocity dispersion at linear order in $\sigma$,
\beq  \langle \vpb_{\sigma,a}^{(1)}(\bk,\eta) \vpb_{ss,b}(\bk',\eta') \rangle=- \int_{\eta_{in}}^\eta ds \; g_{ac}(\eta-s) \left\langle 
 \delta h_{c} \left( \bk, s \right)   \vpb_{ss,b}(\bk',\eta') \right\rangle\,.
\eeq 
Considering the density PS and using a notation analogous to that of Section~\ref{expansionscheme} we can write
\beq
\!\!\!\!\!\!\!\!\!\!\!\! \!\!\!\! \bar P_{{11}} (k,\eta) = \bar P_{{11}}^{lin}(k,\eta)+\bar P_{{ss,11}}^{\rm 1-loop}(k,\eta)+ \bar P_{{ss,11}}^{n\ge {\rm 2-loop}}(k,\eta) + \Delta\bar P_{{11}}^{\delta h}(k,\eta)
{+ {\rm O } \left( \sigma^2 \right) } \,,
 \label{linexp2}
\eeq
where we have split the single stream contribution in its linear plus  1-loop approximation,  and all higher order terms, $\bar  P_{{ss,11}}^{n\ge {\rm 2-loop}}(k,\eta)$. The  $L$-dependence of the various terms is just the trivial $\tilde W[k L]^2$ one, and therefore we have omitted it, or, equivalently, one can imagine taking the limit $L\to 0$.

\begin{figure}[ht!]
\centerline{
\includegraphics[width=0.8\textwidth,angle=0]{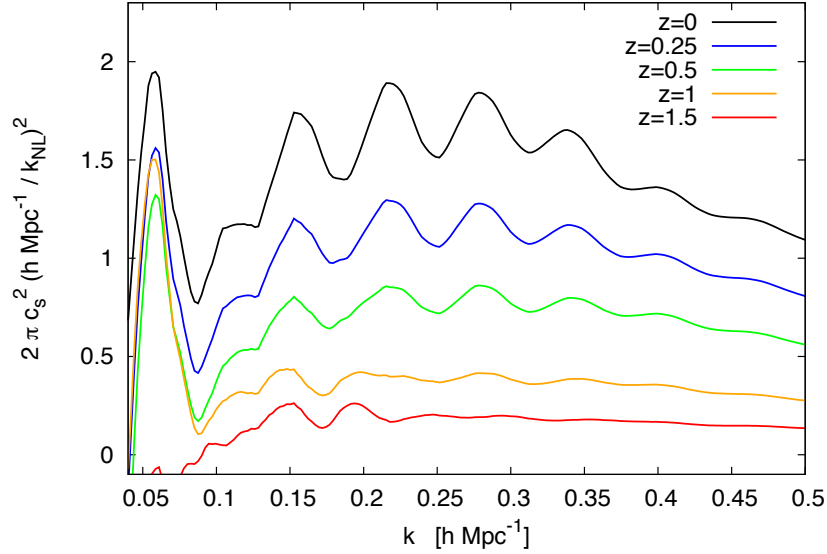}
}
\caption{
Ratio in eq. \re{cs1}  at  $5$ different  redshift in the REF cosmology. 
}
\label{fig:cs2}
\end{figure}

The expression above, which is exact up to $O(\sigma^2)$ corrections, should be compared with the lowest order (1-loop) EFToLSS approximation \cite{Carrasco:2012cv,Carrasco:2013mua}, which, in our notation, {(and removing the $\tW^2 [k L ]$ factor from both sides)} would read (see eq.~(58) of \cite{Carrasco:2013mua}),
\beq
 P_{ {11}} (k,\eta) \simeq   P_{{11}}^{lin}(k,\eta)+  P_{{ss,11}}^{\rm 1-loop}(k,\eta) - 2(2 \pi)\,c_{s(1)}^2 \frac{k^2}{k_{NL}^2}  P^{lin} (k,\eta)\,,
 \label{EFT1l}
\eeq
where we recall that $P^{lin}$ is the linear PS, and where the $k$-independent {parameter} $c_{s(1)}^2$ can be interpreted as an effective speed of sound, to be determined by fitting the PS with N-body simulations. Higher orders in the EFToLSS include higher loops and higher powers of $k/k_{NL}$. In \cite{Carrasco:2013mua}, contributions up to 2-loop orders and $O(k^4/k_{NL}^4)$ have been included, however we will limit our discussion to the comparison with the 1-loop EFToLSS of eq.~\re{EFT1l}, for which the derivation  is straightforward.

Comparing \re{EFT1l} to \re{linexp2} we immediately understand the physical effects contained in the speed of sound term at this order. Namely, $c_{s(1)}^2$ incorporates all the nonlinearities of the pressureless ideal fluid equations beyond 1-loop, plus the effect of the small scale velocity dispersion.

In Figure~ \ref{fig:cs2} we plot the quantity 
\beq
\frac{P_{{ss,11}}^{\rm 1-loop}(k,\eta) -P_{ {11}} (k,\eta) }{2 \, k^2 P^l \left( k , \eta \right) } \;, 
\label{cs1}
\eeq
which, according to eq.~\re{EFT1l}, should correspond to the $k$-independent quantity $\frac{2 \pi c_{s(1)}^2}{  k_{NL}^2}$.  In ref.~\cite{Carrasco:2013mua} the speed of sound was determined by fitting eq.~\re{EFT1l} to the nonlinear PS in the $k \sim 0.15 - 0.25$ h Mpc$^{-1}$ range, obtaining in this way $2 \pi c_{s(1)}^2 \, \left( \frac{ \hmpc }{  k_{NL} } \right)^2= 1.62\pm 0.03 $ for $z=0$.  As we see,  the central value  is consistent with the $z=0$ curve of the REF cosmology  shown in Figure~ \ref{fig:cs2}.  However the residual BAO fluctuations -- coming from the difference between the 1-loop PS and the nonlinear PS-- are quite strong, and the difference between the lines in Figure~\ref{fig:cs2} and a constant  gives a measure of the error made in the measurement of $c_{s(1)}^2$ by ordering the corrections by simple power counting in $k/k_{NL}$ when the PS is not a simple power-law, but has scale dependent features such as BAO's. 
This error propagates, in a reduced manner proportional to the size of the correction, to the final reconstruction of the PS.  In  \cite{Senatore:2014via} these residual fluctuations were reduced by resumming the effect of particle displacements at all orders in SPT.

\begin{figure}[ht!]
\centerline{
\includegraphics[width=0.8\textwidth,angle=0]{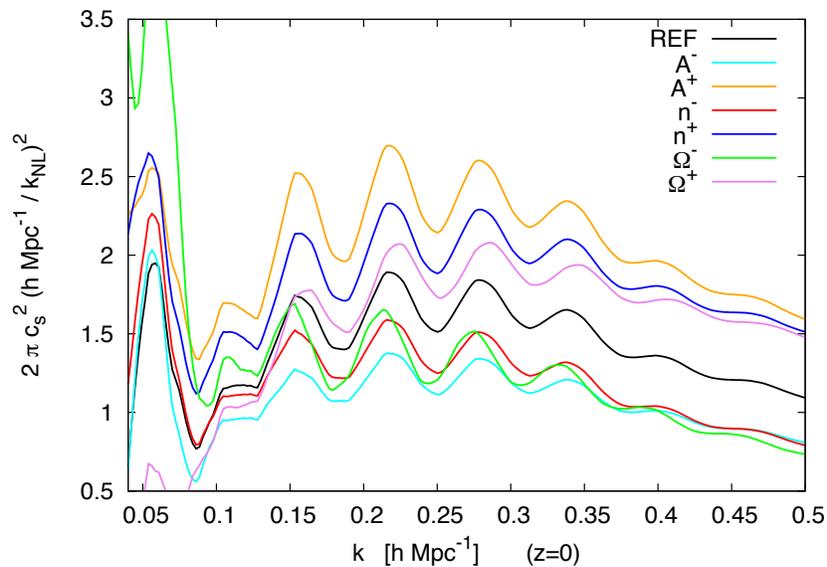}
}
\caption{
Dependence of the sound speed on the  cosmologies in Table~\ref{tab_sims}. The curves show the ratio 
 in eq. \re{cs1} for these cosmologies at  $z=0$. 
}
\label{fig:cs-cosmo}
\end{figure}

Moreover, the sound speed defined according to eq. \re{EFT1l} exhibits a significant cosmology dependence. We show this in Figure~ \ref{fig:cs-cosmo}, where we plot the ratio \re{cs1} at redshift $z=0$  for the cosmologies in Table~\ref{tab_sims}. If we adopt the prescription of \cite{Carrasco:2013mua} that consists in determining the sound speed from the average of this result in the  $k \sim 0.15 - 0.25$ h Mpc$^{-1}$ range we obtain the numerical values reported in Table~\ref{tab:sound}.

\begin{table}
\centering
\begin{tabular}{|l|l|l|l|l|l|l|l|} \hline 
Cosmology & ${\rm REF} $ & $A_s^-$ &  $A_s^+$ &  $n_s^-$ &  $n_s^+$ &  $\Omega_s^-$ &  $\Omega_s^+$    \\ \hline
 ``Average'' $\;\;2 \pi c_{s(1)}^2 \, \left( \frac{ \hmpc }{  k_{NL} } \right)^2$ & 
1.64 & 1.21 & 2.34 & 1.39 & 2.02  & 1.39 & 1.78   \\ \hline
\end{tabular}
\caption{Cosmology dependence of the rescaled sound speed, defined as in \cite{Carrasco:2013mua} from the average of eq. \re{cs1} in the  $k \sim 0.15 - 0.25$ h Mpc$^{-1}$ range, at  $z=0$. 
}
\label{tab:sound}
\end{table}

In the approach followed in this paper, on the other hand, we do not send the cutoff $L\to 0$, but we keep it fixed at a scale such that the 1-loop expression holds for the scales $k\la 1/L$ and we use the N-body simulations to replace the 1-loop contributions for $k\ga 1/L$ with the fully non-perturbative ones. The scale dependence of the correction to the 1-loop result is therefore taken into account more carefully. In the previous sections we have discussed how, once this has been measured for a given redshift and a given cosmology, it can be used for reconstructing the full PS at different redshifts and for different cosmologies.

\begin{figure}[ht!]
\centerline{
\includegraphics[width=0.8\textwidth,angle=0]{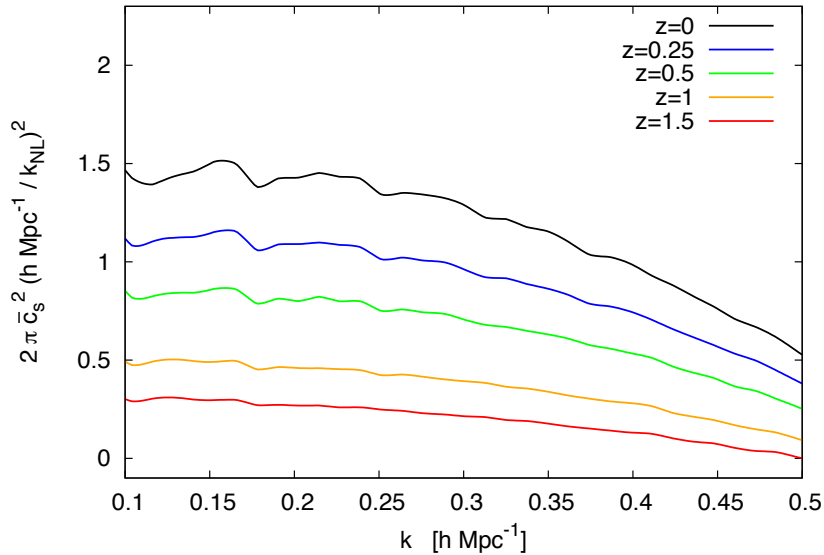}
}
\caption{
Sound speed defined according to our eq. \re{cs-us}  at  $5$ different  redshift in the REF cosmology. We notice that the $k-$depence due to the BAO oscillations is significantly less pronounced than the one obtained from the prescription \re{cs1}, and shown in Figure \ref{fig:cs2}. 
}
\label{fig:cs2-us}
\end{figure}

Our approach suggests a definition of effective sound speed different from \re{cs1}, namely,
\beqra
&&\!\!\! \!\!\! \!\!\! \!\!\! \!\!\! \!\!\! \!\!\! \!\!\! 2 \pi \bar c_s^2 \frac{k^2}{k_{NL}^2} \equiv \frac{\int_{\etain}^\eta ds \,g_{12}(\eta-s) \langle \delta h_2(\bk,s) \vpb_1(\bk^\prime, \eta) \rangle^\prime}{ \langle \vpb_1(\bk,s) \vpb_1(\bk^\prime, \eta) \rangle^\prime}\nonumber\\
&&= -\frac{1}{2} \left( \frac{\Delta \bar P_{11}^{h,{\rm N-body}} (k,\eta) }{ \bar P_{11}^{{\rm N-body}} (k,\eta) }-\frac{ \Delta \bar P_{ss,11}^{h,{\rm 1-loop}} (k,\eta)}{\bar P_{ss,11}^{{\rm 1-loop}} (k,\eta)}\right)+O({\mathrm{2-loop}})\,.
\label{cs-us}
\eeqra
The first line in the above definition relates directly the effective speed of sound to the non-perfect fluid effects, that is, it exactly vanishes in the single stream approximation, and, moreover, it is  exactly independent on the cutoff $L$. This is not true any more by truncating the PT expansion of $h_{ss,2}$ at a finite order, as we do at the second line.  Still, even at this order, the scale dependence is greatly reduced with respect to that of the expression in eq.~\re{cs1}, since taking ratios N-body/N-body and PT/PT cancels the residual BAO's much more efficiently. This is explicitly shown in Figure \ref{fig:cs2-us}. We see however a residual scale dependence at the highest momenta shown, which is to be interpreted as the effect of neglecting $2-$loop contributions $\propto k^4$  in this computation.

\section{Conclusions}
\label{fine}
In this paper we have shown that the lowest order approximation in the CGPT approach, namely, a 1-loop computation for the IR modes and the measurement of only one cross-correlator between the UV source term and the density field, is able to reproduce the nonlinear PS at the percent level up to $k\la 0.4\;\hmpc$. The cross-correlator appears in a time integral, eq.~\re{ccint}, which would require measuring it by cross-correlating different N-body snapshots. Instead, we approximate the time dependence with the SPT-inspired Ansatz \re{alpha}, which allows an analytical evaluation of the integral. This introduces a free parameter, $\alpha(\eta)$, which can be measured either by fitting  the reconstructed PS to the nonlinear one, or by direct measurement of the scaling with time of the cross-correlator in the IR regime. The results from the two approaches give compatible results, which corroborates the Ansatz \re{alpha}.

Once the cross-correlator and $\alpha(\eta)$ have been measured in a given cosmology, they can be used for a different one by using eq.~\re{P-reconstructed3}. Therefore we do not need to run an N-body simulation for each cosmology. Most of the cosmology dependence is encoded in the 1-loop SPT quantity $ \Delta \bar P_{ss,11,i}^{h,{\rm 1-loop}} (k,\eta) $. The residual dependence in the ${\cal C}_i(\eta)$ factor is a very shallow one, as we read from the large error bars in Table~\ref{tab:CGPT} , and, at least for the cosmologies we considered, it can be neglected by setting ${\cal C}_i(\eta)={\cal C}_{REF}(\eta)$. The same holds for the time dependence (the slope) of the $\alpha_i$ functions, as we discussed in Section~\ref{sec:time}.

In any case, a very robust feature of the UV contribution emerges from this investigation, namely the scale-independent ratios of Figure~\ref{fig:simple_ratios} between the source-density  cross-correlators computed for different cosmologies. This shows that a single scale-independent parameter is enough to describe the cosmology change. In our approach, we have shown that this parameter can be mostly reproduced by a proper SPT computation,  Figure \ref{fig:Ri_ratios},  but one could also take the agnostic view of simply marginalizing over it in a parameter estimation procedure. Due to the scale independence, this will probably have small impact on extracting cosmological parameters from BAO observations. 

It will be very important to extend the analysis of this paper to an enlarged set of cosmological models: massive neutrinos, modifications of General Relativity, clustering Dark Energy, and so on, to investigate to what extent the scale independence of Figure~\ref{fig:simple_ratios} holds also for these less standard scenarios. An analytic understanding of the behavior of these UV sources, {\it e.g.} in the halo model, would be precious in order to better understand the physical reason for the scale independent ratios (which is however already in agreement with SPT) and to predict the time-dependence of the $\alpha_i(\eta)$ exponents.

On a more theoretical side, a next order computation (2-loop, $O(\sigma^2)$) would probably not improve too much the results in the BAO region, but it would provide useful information on issues such as the smallest scales which can be described by such method, and would mitigate further the cutoff dependence of the results.

\section*{Acknowledgments}

We thank G. Mangano and L. Hui for discussions. M. Peloso would like to thank the University of Padova, Dipartimento di Fisica e Astronomia, and the INFN, Sezione di Padova, for their friendly hospitality and for partial support during the earlier stages of this work. M. Pietroni would like to thank the Institute for Theoretical Physics of the University of Heidelberg for their friendly hospitality and for partial support during the later stages of this work. A. Manzotti acknowledges partial support from the Kavli Institute for Cosmological Physics at the University of Chicago through grants NSF PHY-1125897 and an endowment from the Kavli Foundation and its founder Fred Kavli.  M. Peloso  acknowledges partial support from the DOE grant DE-SC0011842  at the University of Minnesota.    M. Pietroni acknowledges partial support from the  European Union FP7  ITN INVISIBLES (Marie Curie Actions, PITN- GA-2011- 289442). M. Viel and F. Villaescusa-Navarro are supported by the ERC StG "cosmoIGM" and by INFN IS PD51 "INDARK" grants.

\appendix
\section{Check of the $L$-factorization in the 1-loop density PS in   single stream approximation}
\label{check-factorization}

The filtering operation commutes with the expansion in the microscopic velocity dispersions. A consequence of this are the  identities written in eq. \re{filts} and in the following line of the main text. Each of these identities is actually valid order by order in $\vp^{lin}$. Therefore, the simplest nontrivial identity gives rise to the expression 
\begin{equation}
\langle \vpb_{ss,1} \left( \bk, \eta \right)  \vpb_{ss,1} \left( \bk', \eta \right) \rangle_{\rm 1-loop}' = {\tilde W}^2 \left( k L \right) P_{11}^{\rm 1-loop} \left( k , \eta \right) \,. 
\label{P1loop-Lindep} 
\end{equation} 
In this appendix we explicitly verify this relation, as a check on our expansion scheme. 

Our starting point is the system \re{eom} for the fields in single stream approximation. Is is convenient to define
\begin{equation}
\vpb_{ss,a} \left( \bk, \eta \right) = \vpb_a^{(h=0)} \left( \bk, \eta \right) + \vpb_a^{(h)} \left( \bk, \eta \right) \;\;,\;\; 
\label{phi-deco}
\end{equation}
where 
\beq
\vpb_a^{(h)}(\bk,\eta)\equiv -\int_{\eta_{\rm in}}^\eta ds \, g_{ab}(\eta-s) h_{ss,b}(\bk,s)\,. 
\label{phi-h-def}
\eeq

To obtain the 1-loop PS, we need to expand  \footnote{We stress that the order in each term refers to the number of linear fields contained in it. Contrary, for example to eq. (\ref{expvpb}), these expressions are not series expansions in the microscopic $\sigma$. All this appendix assumes single stream approximation, $\sigma =0$). }
\beqra
\vpb_a^{(h=0)}&=&\vpb_a^{(h=0,1)}+\vpb_a^{(h=0,2)}+\vpb_a^{(h=0,3)}+{\rm O } \left( \left( \vp^{lin} \right)^4 \right) 
\,, \nonumber\\  
\vpb_a^{(h)}&=&\vpb_a^{(h,2)}+\vpb_a^{(h,3)}+{\rm O } \left( \left( \vp^{lin} \right)^4 \right) \,,  
\label{exp-app}
\eeqra
where we keep into account that $h_{ss}$ is at least quadratic in the linear fields. Indeed, the terms in the expansion in $\vpb_a^{(h)}$ can be immediately obtained by expanding the source $h_{ss}$ in \re{phi-h-def}, and,  expanding \re{hsumss} and \re{J1J2ss}, we see that 
\begin{eqnarray}
J_{1,ss}^i &  =  & \langle \delta_{ss} \, \partial_i \phi_{ss} \rangle - {\bar \delta}_{ss} \langle \partial_i \phi_{ss} \rangle - {\bar \delta}_{ss}  \langle \delta \partial_i \phi_{ss} \rangle + {\bar \delta}_{ss}^2 \langle \partial_i \phi_{ss} \rangle + {\rm O } \left( \delta_{ss}^3 \phi_{ss} \right) \;, \nonumber\\ 
J_{\sigma,ss}^i & = & \partial_k \langle v_{ss}^i v_{ss}^k \rangle - \partial_k \left( {\bar v}_{ss}^i {\bar v}_{ss}^k \right) + \partial_k \langle \delta_{ss} v_{ss}^i v_{ss}^k \rangle  - {\bar \delta}_{ss}  \partial_k \langle v_{ss}^i v_{ss}^k \rangle -  {\bar v}_{ss}^i {\bar v}_{ss}^k \partial_k {\bar \delta}_{ss} \nonumber\\ 
& &  + {\rm O } \left( \delta_{ss}^2 v_{ss}^2 \right) \;. 
\label{J-pert}
\end{eqnarray} 

To obtain  $\vpb_a^{(h=0)} $ we instead insert \re{phi-deco} and \re{phi-h-def} into the system \re{eom}. We obtain that the field  $\vpb_a^{(h=0)} $ satisfies   
\beqra
&&\!\!\!\!(\delta_{ab}\partial_\eta +\Omega_{ab})\vpb_b^{(h=0)}(\bk,\eta)= I_{\bk,\bq_1,\bq_2}  e^\eta\gamma_{abc}(\bq_1,\bq_2) \vpb_b^{(h=0)}(\bq_1,\eta)\vpb_c^{(h=0)}(\bq_2,\eta) \nonumber\\
&& \qquad\qquad\qquad\qquad \qquad +2\, I_{\bk,\bq_1,\bq_2}  e^\eta\gamma_{abc}(\bq_1,\bq_2) \vpb_b^{(h=0)}(\bq_1,\eta)\vpb_c^{(h)}(\bq_2,\eta) \nonumber\\
&&\qquad\qquad \qquad\qquad  \qquad- I_{\bk,\bq_1,\bq_2}  D_\omega^{ab}(\bq_1,\bq_2)\vpb_b^{(h=0)}(\bq_2,\eta) + {\rm O } \left( \left( \vp^{lin} \right)^4 \right) \,.\nonumber\\
\label{eom3}
\eeqra

The last explicit  term in this expression originates from the last term in \re{eom}, with only the leading ${\rm O } \left( \left( \vp^{lin} \right)^2 \right) $ contribution retained in the expression  \re{sources_vort} for the vorticity: 
\beqra
\!\!\!\!  \!\!\!\! \!\!\!\! \!\!\!\!  \!\!\!\! \!\!\!\! \!\!\!\!  \!\!\!\! D_\omega^{ab}(\bq_1,\bq_2) & = & \left(
\begin{array}{cc} 
1 & 0 \\
0 & \frac{q_2^2+2\,\bq_1\cdot\bq_2}{q_2^2} \\
\end{array}
\right)_{ab} \, 
 e^{2 \eta}  I_{\bq_1,\bp_1,\bp_2} \left(\frac{\bq_2\cdot\bq_1\, \bq_1\cdot\bp_2}{q_1^2p_2^2}-\frac{\bq_2\cdot\bp_2}{p_2^2}\right)\nonumber\\
&& \times \left[ 
{\tilde W} \left( q_1 L \right) - {\tilde W} \left( p_1 L \right)  {\tilde W} \left( p_2 L \right) 
\right] \vp^{lin} \left( \bp_1 \right) \vp^{lin} \left( \bp_2 \right) 
+ {\rm O } \left( \left( \vp^{lin} \right)^3 \right)\,.\nonumber\\
\label{w-pert}
\eeqra
We also note that the $h_a^\omega$ term in \re{eom} contributes to \re{eom3} only at ${\rm O } \left( \left( \vp^{lin} \right)^4 \right)$ or higher, and the vorticity is at least quadratic in the linear fields. Therefore, this term is irrelevant for the present computation. 

We now proceed by explicitly computing the terms in \re{exp-app}, and then the 1-loop power spectrum. We evaluate the  1-loop mode-mode coupling contribution, also identified as $P^{(22)}_{PT}$ in the standard PT terminology, separately from the  $P^{(13)}_{PT}$ contribution, and we verify that they separately satisfy the identity \re{P1loop-Lindep}. We then conclude this appendix by collecting the results that give the 1-loop PS for the density fields entering in eq. \re{Pr1l}. 

\subsection{ $P^{(22)}_{PT}$ contribution} 
\label{app-P22}

We now compute 
\beqra
\langle \vpb_a^{(2)}(\bk,\eta) \vpb_b^{(2)}(\bk',\eta)\rangle = 
{\cal C}^{00}_{ab} + {\cal C}^{0h}_{ab} + {\cal C}^{hh}_{ab} \;, 
\label{P22-app}
\eeqra
where 
\beqra
{\cal C}^{00}_{ab} & \equiv & \langle \vpb_a^{(h=0,2)}(\bk,\eta) \vpb_b^{(h=0,2)}(\bk',\eta)\rangle \,, \nonumber\\ 
{\cal C}^{0h}_{ab} & \equiv & \langle \vpb_a^{(h=0,2)}(\bk,\eta) \vpb_b^{(h,2)}(\bk',\eta)\rangle + 
 \langle \vpb_a^{(h,2)}(\bk,\eta) \vpb_b^{(h=0,2)}(\bk',\eta)\rangle \,, \nonumber\\ 
{\cal C}^{hh}_{ab} & \equiv & \langle \vpb_a^{(h,2)}(\bk,\eta) \vpb_b^{(h,2)}(\bk',\eta)\rangle \,.  
\eeqra

Only the first line of \re{eom3} and the leading expressions in \re{J-pert} contribute to the second order fields. We obtain
\beqra
&& \!\!  \!\!  \!\! \!\! \vpb_a^{(h=0,2)}(\bk,\eta)= \int^\eta ds\,g_{ac}(\eta-s) e^s \,  I_{\bk,\bq_1,\bq_2}  \nonumber\\
&& \qquad \times\; \gamma_{cde}(k,q_1,q_2)u_d u_e \,\tW[q_1 L] \tW[q_2 L] \,
\vp^{lin} (\bq_1)\vp^{lin} (\bq_2) \,, 
\label{phi2}
\eeqra
and
\beqra
&& \!\!  \!\!  \!\! \!\! \vpb_a^{(h,2)}(\bk,\eta) = \int^\eta ds\,g_{ac}(\eta-s) e^s \,   I_{\bk,\bq_1,\bq_2} \nonumber\\
&& \qquad  \times \; \gamma^h_{cde}(k,q_1,q_2)u_d u_e \,(\tW[kL]-\tW[q_1 L] \tW[q_2 L]) \,
\vp^{lin} (\bq_1)\vp^{lin} (\bq_2) \,, 
\label{phi2h}
\eeqra
where we have defined a new ``vertex" function $\gamma^h_{cde}$, whose only non vanishing elements are,
\beqra
&&\gamma^h_{211}(k,q_1,q_2)=\frac{3}{4}\frac{\Omega_m}{f^2} \left(\frac{\bk\cdot\bq_1}{q_1^2}+\frac{\bk\cdot \bq_2}{q_2^2}\right)\,,
\nonumber \\
&&\gamma^h_{222}(k,q_1,q_2)= \frac{\bk\cdot\bq_1 \bk\cdot\bq_2}{q_1^2 q_2^2}\,.
\eeqra

We then obtain
\beqra
&&\!\!\!\!\!\!\!\! \!\!\!\!\!\!\!\! {\cal C}^{\left(00,0h,hh\right)}_{ab} =  2 \; \left( 2 \pi \right)^3\, \delta_D(\bk+\bk')  I_{\bk,\bq,\bp} P^l \left( q \right) P^l \left( p \right) \nonumber\\
&&  \times  \int^\eta ds\,g_{ac}(\eta-s)e^s\int^\eta ds'\,g_{bd}(\eta-s')e^{s'}  u_e u_f u_g u_h {\cal A}_{cefdgh}^{\left(00,0h,hh\right)} \;, 
\eeqra
where $P^l$ is the power spectrum of the linear fields, and where 
\beqra
{\cal A}_{cefdgh}^{\left(00\right)} &=& \gamma_{cef}(k,p,q)\gamma_{dgh}(k,p,q) \tW[q L]^2 \tW[p L]^2 \;, \nonumber\\ 
{\cal A}_{cefdgh}^{\left(0h\right)} &=& 
\left[ \gamma_{cef}(k,p,q)\gamma^h_{dgh}(k,p,q)+ \gamma^h_{cef}(k,p,q)\gamma_{dgh}(k,p,q) \right] \nonumber\\
&&\qquad\qquad\qquad \times \; \Big(\tW[k L]- \tW[p L]\tW[q L]\Big)\tW[p L]\tW[q L] \;, \nonumber\\ 
{\cal A}_{cefdgh}^{\left(hh\right)} &=& \gamma^h_{cef}(k,p,q)\gamma^h_{dgh}(k,p,q) \Big( \tW[k L]- \tW[p L]\tW[q L] \Big)^2  \,.
\eeqra

We now add the three contributions as in \re{P22-app}, and set $a=b=1$ for the mode coupling part of the PS of the density field. By using the explicit expressions for the vertices we obtain 
\beq
\langle \vpb_1^{(2)}(\bk,\eta) \vpb_1^{(2)}(\bk',\eta)\rangle = \left( 2 \pi \right)^3\delta_D(\bk+\bk') \tW[k L]^2 P^{(22)}_{PT}(k,\eta)\,.
\label{chk1}
\eeq
showing that the mode coupling contribution to the density PS factors out the filter function, in agreement with 
 \re{P1loop-Lindep}.

\subsection{ $P^{(13)}_{PT}$ contribution }

These contribution to the PS is obtained by correlating 
\beq
\vpb_a^{(h=0,1)} \left( \bk, \eta \right) = \tW \left[ k L \right] \, \phi^l \left( \bk \right) u_a \,, 
\eeq
with the third order terms in the expansion \re{exp-app}. 

Let us first discuss the terms with $\vpb_a^{(h=0,3)}$  in the correlator. One contribution to  $\vpb_a^{(h=0,3)}$ 
is obtained from a diagram with two vertices coming both from  the first line in eq. \re{eom3}  (in practice, we use eq. \re{phi2} alternatively for the two fields at the first term at RHS of eq. \re{eom3}). This gives 
\beqra
&&\!\!\!\!\!\!\!\! \!\!\!\!\!\!\!\! 
\langle \vpb_a^{(h=0,3)}(\bk,\eta) \vpb_b^{(h=0,1)}(\bk',\eta)\rangle_{\gamma\gamma} + (a\leftrightarrow b) =  4 \;\left(2 \pi\right )^3 \delta_D(\bk+\bk')  \nonumber\\
&& \int^\eta ds \int^s ds' e^{s+s'} g_{ac}(\eta-s)P^l(k)  I_{\bk,\bq,\bp} P^l(q)u_e u_g u_h u_b\times  \nonumber\\
&& \qquad \gamma_{cde}(k,p,q)  g_{df}(s-s')\gamma_{fgh}(p,q,k) \tW[q L]^2 \tW[k L]^2
+ (a\leftrightarrow b) \,.
\label{p131}
\eeqra

Another contribution  to  $\vpb_a^{(h=0,3)}$ is obtained by using eq.~\re{phi2h} in the second line of \re{eom3}. This gives   
\beqra
&&\!\!\!\!\!\!\!\! \!\!\!\!\!\!\!\! 
\langle \vpb_a^{(h=0,3)}(\bk,\eta) \vpb_b^{(h=0,1)}(\bk',\eta)\rangle_{\gamma\gamma^h} + (a\leftrightarrow b) = 
 4 \; \left( 2 \pi \right)^3 \delta_D(\bk+\bk') \nonumber\\
&&  \int^\eta ds \int^s ds' e^{s+s'} g_{ac}(\eta-s)P^l(k)  I_{\bk,\bq,\bp} P^l(q) u_e u_g u_h u_b \times  \nonumber\\
&& \gamma_{cde}(k,p,q) g_{df}(s-s')\gamma^h_{fgh}(p,q,k) \Big(\tW[k L]\tW[q L] \tW[p L]- \tW[q L]^2 \tW[k L]^2\Big) \nonumber\\
&&+ (a\leftrightarrow b) \,.
\eeqra
The third and final contribution to $\vpb_a^{(h=0,3)}$ comes from the third line of \re{eom3}, the vorticity term, in which we use \re{w-pert}. This gives 
\beqra
&&\!\!\!\!\!\!\!\! \!\!\!\!\!\!\!\! 
\langle \vpb_a^{(h=0,3)}(\bk,\eta) \vpb_b^{(h=0,1)}(\bk',\eta)\rangle_{h0\omega} + (a\leftrightarrow b) =  -  \left( 2 \pi \right)^3 \delta_D(\bk+\bk') \nonumber\\
&&  \!\!\!\!\!\!\!\!  \int^\eta ds\,e^{2s}  P^l(k)  I_{\bk,\bq,\bp} P^l(q) 
\left[ g_{a1} \left( \eta-s \right) u_1 +  \frac{q^2+2\,\bp\cdot\bq}{q^2} \, g_{a2} \left( \eta - s \right) u_2 \right] 
 u_b \, \times \nonumber \\
&& \left[1-\frac{(\bq\cdot\bp)^2}{p^2q^2} + \frac{\bq\cdot \bp \;\bk\cdot\bp}{k^2p^2} - \frac{\bk\cdot \bq}{k^2}\right]\Big(\tW[k L]\tW[q L] \tW[p L]- \tW[k L]^2 \tW[q L]^2\Big)\,\nonumber\\
&&+ (a\leftrightarrow b)\,. 
\eeqra

The remaining contributions to $P^{13}_{PT}$ have  $\vpb_a^{(h,3)}$  in the correlator. 

One of these contribution originates from the PT expansion (namely, eq. \re{phi2}) of the quadratic terms in the sources \re{J-pert}. It gives
\beqra
&&\!\!\!\!\!\!\!\! \!\!\!\!\!\!\!\! 
\langle \vpb_a^{(h,3)}(\bk,\eta) \vpb_b^{(h=0,1)}(\bk',\eta)\rangle_{\gamma^h\gamma} + (a\leftrightarrow b) = 
 4 \;\left(2 \pi\right)^3 \delta_D(\bk+\bk') 
\nonumber\\
&&\!\!\!\!\!\!\!\!  \int^\eta ds \int^s ds' e^{s+s'} g_{ac}(\eta-s)P(k)  I_{\bk,\bq,\bp} P(q)u_e u_g u_h u_b\times  \nonumber\\
&&  \gamma^h_{cde}(k,p,q)  g_{df}(s-s')\gamma_{fgh}(p,q,k) \Big( \tW[k L]^2- \tW[k L]^2 \tW[q L]^2\Big)\,\nonumber\\
&&+ (a\leftrightarrow b)\,.
\label{p13-mid}
\eeqra
(We  recall that the quadratic terms already factored a vertex $\gamma^h$, cf. eq. \re{phi2h}, while the vertex $\gamma$ originates from the PT expansion). 

Taking into account that the $- \vpb^2$ terms in the sources can be expanded also using \re{phi2h}, we have also the contribution
\beqra
&&\!\!\!\!\!\!\!\! \!\!\!\!\!\!\!\! 
\langle \vpb_a^{(h,3)}(\bk,\eta) \vpb_b^{(h=0,1)}(\bk',\eta)\rangle_{\gamma^h\gamma^h} + (a\leftrightarrow b) = 
 - 4 \; \left(2 \pi\right)^3 \delta_D(\bk+\bk') 
\nonumber\\
&&\!\!\!\!\!\!\!\!    \int^\eta ds \int^s ds' e^{s+s'} g_{ac}(\eta-s)P^l(k)  I_{\bk,\bq,\bp} P^l(q)u_e u_g u_h u_b\times  \nonumber\\
&& \gamma^h_{cde}(k,p,q)  g_{df}(s-s')\gamma^h_{fgh}(p,q,k) \Big( \tW[k L]\tW[q L]\tW[p L]- \tW[k L]^2 \tW[q L]^2\Big)\,\nonumber\\
&&+ (a\leftrightarrow b)\,.
\eeqra
Then, we have the contribution from the vorticity component of $\bar v^i$ in $J_\sigma^i$, which gives
\beqra
&&\!\!\!\!\!\!\!\! \!\!\!\!\!\!\!\! 
\langle \vpb_a^{(h,3)}(\bk,\eta) \vpb_b^{(h=0,1)}(\bk',\eta)\rangle_{h1\omega} + (a\leftrightarrow b) = 
2 \; \left(2 \pi\right)^3 \delta_D(\bk+\bk')  \nonumber\\
&&\!\!\!\!\!\!\!\!   \int^\eta ds e^{2s}g_{a2}(\eta-s) u_b P^l(k) I_{\bk,\bq,\bp} P^l(q) 
\, \frac{\bk \cdot \bq }{q^2} \frac{k^2-q^2}{p^2} \left[ 1 - \frac{\left( \bk \cdot \bq \right)^2}{k^2 q^2} \right] \nonumber\\ 
&&\Big( \tW[k L]\tW[q L]\tW[p L]- \tW[k L]^2 \tW[q L]^2\Big) + (a\leftrightarrow b)\,.
\eeqra

Finally, the cubic terms in \re{J-pert} give 
\beqra
&&\!\!\!\!\!\!\!\! \!\!\!\!\!\!\!\! 
\langle \vpb_a^{(h,3)}(\bk,\eta) \vpb_b^{(h=0,1)}(\bk',\eta)\rangle_{h^{(3)}} + (a\leftrightarrow b) = 
\left(2 \pi\right)^3 \delta_D(\bk+\bk')  \nonumber\\
&&\!\!\!\!\!\!\!\!  \int^\eta ds e^{2s}g_{a2}(\eta-s) u_b P^l(k) I_{\bk,\bq,\bp} P^l(q) \times\nonumber\\
&&\Bigg\{\frac{3}{2} \frac{\Omega_m}{f^2} \left(1-\frac{\bk\cdot\bq}{q^2}\right)\left(\tW[k L]^2 \tW[q L]^2-\tW[k L] \tW[q L]\tW[p L] \right)\nonumber\\
&&- \frac{(\bk\cdot\bq)^2}{q^4}\tW[k L]^2 +\left(1+\frac{(\bk\cdot\bq)^2}{k^2 q^2}+\frac{(\bk\cdot\bq)^2}{q^4}\right)\tW[k L]^2\tW[q L]^2 \nonumber\\
&&+\left(\frac{\bq\cdot\bp}{q^2}+\frac{\bk\cdot\bq\,\bk\cdot\bp}{k^2 q^2}\right)\tW[k L] \tW[q L] \tW[p L]\Bigg\}+ (a\leftrightarrow b)\,.
\label{p13l}
\eeqra
Setting again $a=b=1$, the sum of eqs.~\re{p131} to \re{p13l} gives 
\beq
2 \langle \vpb_1^{(3)}(\bk,\eta) \vpb_1^{(1)}(\bk',\eta)\rangle  = \left(2 \pi\right)^3 \delta_D(\bk+\bk') \tW[k L]^2 P^{(13)}_{PT}(k,\eta)\,,
\eeq
which, together with the result \re{chk1} completes our check that the only cutoff dependence of the 1-loop result for the density density PS is in the overall factor $\tW[k L]^2$, in agreement with eq. \re{P1loop-Lindep} of the main text.

\subsection{Explicit expressions for the 1-loop PS}
\label{app-explicit}

We now provide the explicit expressions for the 1-loop power spectra used in the main text, cf. eq. \re{P-reconstructed}. 

We have 
\begin{equation}
\langle \vpb_{ss,1} \left( \bk , \eta \right)  \vpb_{ss,1} \left( \bk' , \eta \right) \rangle' \equiv  \bar P_{11,ss} ^{\rm 1-loop } =  \bar P_{11,ss} ^{(22)} +  \bar P_{11,ss} ^{(13)} \,. 
\end{equation}
The mode coupling term evaluated in  \ref{app-P22} acquires the explicit final expression
\begin{eqnarray}
& & \!\!\!\!\!\!\!\! \!\!\!\!\!\!\!\! \bar P_{11,ss} ^{(22)} 
 = \tW^2\left[ k \, L \right] \, \frac{1}{\left( 2 \pi \right)^3}  {\rm e}^{2 \eta} \frac{k^3 \pi}{49} \nonumber\\ 
& &  \int_0^\infty d r  P^l \left( k r \right) \int_{-1}^1 d x \,  P^l \left( k \sqrt{1+r^2 - 2 r x} \right) 
 \frac{\left( 3 r + 7 x - 10 r x^2 \right)^2}{\left( 1 + r^2 - 2 r x \right)^2} \;,   
\end{eqnarray}
where we recall that $P^l$ is the PS of the linear $\vp^{lin}$ fields. The sum of expressions ~\re{p131} to \re{p13l} gives instead 
\begin{eqnarray}
& &  \!\!\!\!\!\!\!\! \!\!\!\!\!\!\!\! \bar P_{11,ss} ^{(13)} = 
  \tW^2 \left[ k \, L \right] \,  P^l \left( k \right) 
\frac{1}{\left( 2 \pi \right)^3} 2 {\rm e}^{2 \eta} \frac{k^3 \pi}{252} \,   \int_0^\infty d r P^l \left( k r \right) 
 \nonumber\\ 
& &  \left[ \frac{12}{r^2} - 158 + 100 r^2 - 42 r^4 + \frac{3}{r^3}
\left( r^2 - 1 \right)^3 \left( 2 + 7 r^2 \right) {\rm ln } \left\vert \frac{r+1}{r-1} \right\vert \right]  \,. 
\end{eqnarray} 
We remark that these expressions are the standard 1-loop PT results, times the square of the filter function.

We then  have 
\begin{eqnarray}
&& \!\!\!\!\!\!\!\!  \!\!\!\!\!\!\!\!  \!\!\!\!\!\!\!\! 
- \int_{\eta_{\rm in}}^\eta d s g_{1c} \left( \eta - s \right) \left\langle h_{ss,c} \left( \bk , s \right) \vpb_{ss,1} \left( \bk' , \eta \right) \right\rangle'_{\rm 1-loop} + \left( \bk \leftrightarrow \bk' \right) \nonumber\\ 
&& =  2  \left\langle \vpb_1^{(h)} \left( \bk , \eta \right)  \vpb_{ss,1} \left( \bk' , \eta \right) \right\rangle'_{\rm 1-loop} \nonumber\\ 
& & \equiv  \Delta \bar P_{ss,11}^{h,{\rm 1-loop}} (k,\eta) =  \Delta \bar P_{ss,11}^{h,(22)} (k,\eta) +  \Delta \bar P_{ss,11}^{h,(13)} (k,\eta)  
\end{eqnarray}

The mode coupling contribution is given by the $ {\cal C}^{0h}_{ab} $ term in eq. \re{P22-app}, and it evaluates to 

\beqra
&& \!\!\!\!\!\!\!\!  \!\!\!\!\!\!\!\! 
\Delta \bar P_{ss,11}^{h,(22)} (k,\eta)    = 
\frac{k^3 {\rm e}^{2 \eta}}{196 \pi^2 }   \int_0^\infty d r \tW [ k r L ] P^l \left( k r \right) 
 \nonumber\\ 
&& 
\int_{-1}^1 d \xi \,  P^l \left(  k \sqrt{1-2 r \xi + r^2} \right)  
 \frac{\left( 3 r + 7 \xi - 10 r \xi^2 \right)^2}{\left(1-2r\xi+r^2 \right)^2} \tW [ k \sqrt{1-2 r \xi + r^2} L ] 
\nonumber\\ 
&&  \qquad\qquad\qquad\qquad \left( \tW[ k L ] - \tW [ k r L ] \tW [ k \sqrt{1-2 r \xi + r^2} L  ] \right)  \,.  
\eeqra

The other term is obtained by summing the contributions from \re{p13-mid} to \re{p13l}, giving 
\beqra
&& \!\!\!\!\!\!\!\!  \!\!\!\!\!\!\!\! 
\Delta \bar P_{ss,11}^{h,(13)} (k,\eta)    = 
- \frac{3 {\rm e}^{2 \eta} k^3 P^l \left( k \right) \tW \left[ k L \right] }{1512 \pi^2 } \int_0^\infty  d r \, P^l \left( r k \right) \nonumber\\ 
&& \!\!\!\!\!\!\!\!   \Bigg\{ 
\left[ 21 r^4 - 50 r^2 + 79 - \frac{6}{r^2} + \frac{3}{2} \frac{\left( 1 - r^2 \right)^3 \left( 2 + 7 r^2 \right)}{r^3} \, 
  {\rm ln } \left\vert \frac{r+1}{r-1} \right\vert \right] \tW \left[ k L \right]  
\nonumber\\ 
&& \!\!\!\!\!\!\!\!   + \frac{28}{3} \left( 2 - 17 r^2 \right) \tW \left[ k L \right] \tW^2 \left[ k r L \right] 
 + 2 \tW \left[ k r L \right]  \int_{-1}^1 d \xi   \tW \left[ k \sqrt{1+r^2-2 r \xi} L \right] \times \nonumber\\ 
&& \quad  
 \left( r - \xi \right)   \frac{\left( 35+14 \xi^2 \right) r^3 - \xi \left( 77 - 56 \xi^2 \right) r^2 + 5 \left( 13 - 34 \xi^2 \right) r + 77 \xi }{1 + r^2 - 2 r \xi } \Bigg\} \,. 
\eeqra

\subsection{Explicit expression of the N-body correlator in the REF cosmology}

The RHS of eq. \re{P-reconstructed3} is our main result for the density PS of an arbitrary cosmology. For this expression we only need to compute once for all the source in the REF cosmology through an N-body simulation. Specifically, we can rewrite eq.  \re{P-reconstructed3} as 
\beqra
 \bar P_{11,i} (k,\eta) & \simeq & \bar P_{ss,11,i} ^{\rm 1-loop} 
- \left\{ 1+ {\cal C}_i \left( \eta \right) {\cal N} \left( k , \eta \right)   \right\}  \Delta \bar P_{ss,11,i}^{h,{\rm 1-loop}} (k,\eta) \,, \nonumber\\ 
{\cal N} \left( k, \eta \right) & \equiv   &   \left( \frac{ D_{\rm REF} \left( \eta_* \right)}{D_{\rm REF} \left( \eta \right) } \right)^{\alpha_{\rm REF} \left( \eta \right) - 2 }  \frac{ \langle h_2 \left( \bk, \eta \right)   \vpb_{1}(\bk',\eta) \rangle_{\rm REF}' }{ \Delta \bar P_{ss,11,{\rm REF}}^{h,{\rm 1-loop}} (k,\eta)} \,,  
\label{cal-N} 
\eeqra
where the cosmology-dependent coefficient ${\cal C}_i$ is given in eq. \re{calC}. All the information from the REF cosmology is encoded in the function ${\cal N} \left( k, \eta \right)$. We give the values of this function at various momenta in our grid and at various redshifts in Tables~\ref{tab:N1} and \ref{tab:N2}. These are the values used in Section \ref{sec:noNbody}.

\begin{table}
\centering
\begin{tabular}{|l|l|l|l|l|l|} \hline
$ k  \; [ h \, {\rm Mpc}^{-1} ]  $ & $  z=0 $ & $ z = 0.25 $ & $ z = 0.5 $ & $ z=1 $ & $ z=1.5 $ \\ \hline
0.04189 &  -22.769 &  -21.978 &  -20.886 &  -20.86 &  -20.854 \\ \hline
0.05437 &  -14.654 &  -14.238 &  -13.967 &  -13.922 &  -14.143 \\ \hline
0.06709 &  -11.02 &  -11.071 &  -11.007 &  -10.9 &  -11.228 \\ \hline
0.07906 &  -10.784 &  -10.618 &  -10.404 &  -10.429 &  -10.65 \\ \hline
0.09134 &  -10.657 &  -10.457 &  -10.402 &  -10.299 &  -10.424 \\ \hline
0.1037 &  -9.5307 &  -9.2659 &  -9.1511 &  -9.1649 &  -9.3105 \\ \hline
0.1161 &  -9.5171 &  -9.4227 &  -9.2084 &  -9.2901 &  -9.4307 \\ \hline
0.1284 &  -9.9881 &  -9.8072 &  -9.579 &  -9.6157 &  -9.6923 \\ \hline
0.1406 &  -9.6142 &  -9.3851 &  -9.1766 &  -9.1628 &  -9.1966 \\ \hline
0.1529 &  -9.619 &  -9.2895 &  -9.0581 &  -8.8711 &  -8.9044 \\ \hline
0.1654 &  -9.8505 &  -9.6062 &  -9.3549 &  -9.2146 &  -9.1739 \\ \hline
0.1777 &  -8.7385 &  -8.5163 &  -8.313 &  -8.2408 &  -8.2664 \\ \hline
0.1899 &  -8.5702 &  -8.3233 &  -8.1506 &  -8.0763 &  -8.0655 \\ \hline
0.2021 &  -8.6357 &  -8.3204 &  -8.0134 &  -7.8816 &  -7.8148 \\ \hline
0.2144 &  -8.5725 &  -8.2674 &  -8.0855 &  -7.8846 &  -7.8426 \\ \hline
0.2267 &  -8.7537 &  -8.5019 &  -8.2564 &  -8.1066 &  -7.9504 \\ \hline
0.239 &  -8.3866 &  -8.0742 &  -7.8584 &  -7.6715 &  -7.6011 \\ \hline
0.2513 &  -8.4467 &  -8.0667 &  -7.774 &  -7.5224 &  -7.4443 \\ \hline
0.2636 &  -8.3299 &  -7.9773 &  -7.6964 &  -7.4345 &  -7.2596 \\ \hline
0.2758 &  -7.8928 &  -7.4875 &  -7.1876 &  -6.9159 &  -6.7223 \\ \hline
0.2881 &  -7.8624 &  -7.4591 &  -7.1588 &  -6.8676 &  -6.7369 \\ \hline
0.3004 &  -7.7228 &  -7.2474 &  -6.9001 &  -6.6069 &  -6.4249 \\ \hline
0.3126 &  -7.4547 &  -7.0516 &  -6.7616 &  -6.5887 &  -6.4404 \\ \hline
0.325 &  -7.2651 &  -6.9177 &  -6.6184 &  -6.3522 &  -6.1989 \\ \hline
0.3373 &  -7.169 &  -6.8044 &  -6.5113 &  -6.2678 &  -6.1105 \\ \hline
\end{tabular}
\caption{Table of values for the function ${\cal N} \left( k , \eta \right)$ defined in eq. \re{cal-N}. 
This table is continued by Table~\ref{tab:N2}. 
}
\label{tab:N1}
\end{table}

\begin{table}
\centering
\begin{tabular}{|l|l|l|l|l|l|} \hline
$ k  \; [ h \, {\rm Mpc}^{-1} ]  $ & $  z=0 $ & $ z = 0.25 $ & $ z = 0.5 $ & $ z=1 $ & $ z=1.5 $ \\ \hline
0.3496 &  -6.9646 &  -6.5528 &  -6.2584 &  -6.0055 &  -5.8622 \\ \hline
0.3618 &  -6.8781 &  -6.5329 &  -6.2598 &  -5.9708 &  -5.8223 \\ \hline
0.374 &  -6.438 &  -6.1324 &  -5.8862 &  -5.664 &  -5.5419 \\ \hline
0.3863 &  -6.3107 &  -5.9708 &  -5.697 &  -5.4749 &  -5.3555 \\ \hline
0.3986 &  -6.218 &  -5.8443 &  -5.5441 &  -5.3881 &  -5.2805 \\ \hline
0.4109 &  -5.9557 &  -5.6486 &  -5.4075 &  -5.2213 &  -5.1131 \\ \hline
0.4232 &  -5.8415 &  -5.5234 &  -5.2841 &  -5.0915 &  -4.9649 \\ \hline
0.4355 &  -5.5195 &  -5.2324 &  -5.0133 &  -4.8701 &  -4.777 \\ \hline
0.4478 &  -5.3941 &  -5.1071 &  -4.8755 &  -4.6829 &  -4.6088 \\ \hline
0.46 &  -5.0666 &  -4.8823 &  -4.6988 &  -4.5985 &  -4.4877 \\ \hline
0.4722 &  -5.0114 &  -4.7449 &  -4.5356 &  -4.3721 &  -4.2662 \\ \hline
0.4845 &  -4.8651 &  -4.578 &  -4.3759 &  -4.2946 &  -4.2675 \\ \hline
0.4968 &  -4.5482 &  -4.3485 &  -4.1928 &  -4.0874 &  -4.0197 \\ \hline
0.5091 &  -4.436 &  -4.2369 &  -4.0683 &  -4.0069 &  -3.9645 \\ \hline
\end{tabular}
\caption{Continuation of Table~\ref{tab:N1}. 
}
\label{tab:N2}
\end{table}
\newpage

\section{Choice of the filter scale} \label{choice-L}
\label{filterdep}

In this Appendix we discuss the choice of the scale $L=2 \;\mpch$ in the filter function 
\beq
\tilde W(k\,L) = e^{-\frac{k^2 L^2}{2}}\,,
\label{filter-app}
\eeq
that we have adopted in our numerical evaluations. As mentioned in Section \ref{NBODYSIM}, we use numerical simulations of $n_{part} = 512^3$ particles on a box of $L_{box}=512  \;\mpch$ comoving size. The quantities needed for the computation of the sources \re{J1} and \re{Jsig}, namely, particle number density, mass-weighted velocity, acceleration, and velocity dispersion, eq.~\re{n-v-sigma-bar}, are measured on a grid of step $d= 2 \; \mpch$. In this way we obtain fields with wave numbers up to $k_{max} \simeq \sqrt{3} \pi/d\simeq 2.72\;\hmpc$ in Fourier space, which we consider as our UV fields, that is, the fields without the overbars in eq.~\re{n-v-sigma-bar}, \re{J1} and \re{Jsig}. Then, in order to compute the sources, we filter products of the fluctuations of these UV fields with the gaussian in \re{filter-app}, which damps out wave numbers  with $k  \gg k_{smooth} \simeq  \sqrt{2} /L$. Since the UV fields are defined only up to $k_{max}$, one should have $k_{smooth} \ll k_{max}$, which means, $L \gg \sqrt{2} d/(\sqrt{3} \pi) \simeq 0.52\; \;\mpch$ in our configuration.
For this reason, and also to limit the the upper limit of SPT loop integrations (given by $k_{smooth}$) in a reliable range, we did not consider a scale smaller than $L= 2  \;\mpch$ for our filter function, corresponding to $k_{smooth}\simeq  0.71\; \hmpc$.

We stress, however, that the procedure described above takes into account, through the sources, the UV information up to the Nyquist wave number of the simulation, namely $\sqrt{3} \,\pi \,n_{part}^{1/3}/L_{box}\; \hmpc \simeq 5.44\;\hmpc$, and not only up to $k_{max}$. Indeed, we could have avoided the intermediate step of defining the UV fields up to $k_{max}$, and measure the fields and the sources, filtered at the scale $L$, starting directly from the particles. We followed the described procedure only for practical purposes, since processing the UV fields once they have been measured from the simulation (e.g., computing the sources and the cross-correlators) can be done on a normal laptop.

We performed a comparison between $L= 2  \;\mpch$ and  $L= 4  \;\mpch$ and we obtained a much better reconstruction in the former case. 

\begin{figure*}[ht!]
\centerline{
\includegraphics[width=0.5\textwidth,angle=0]{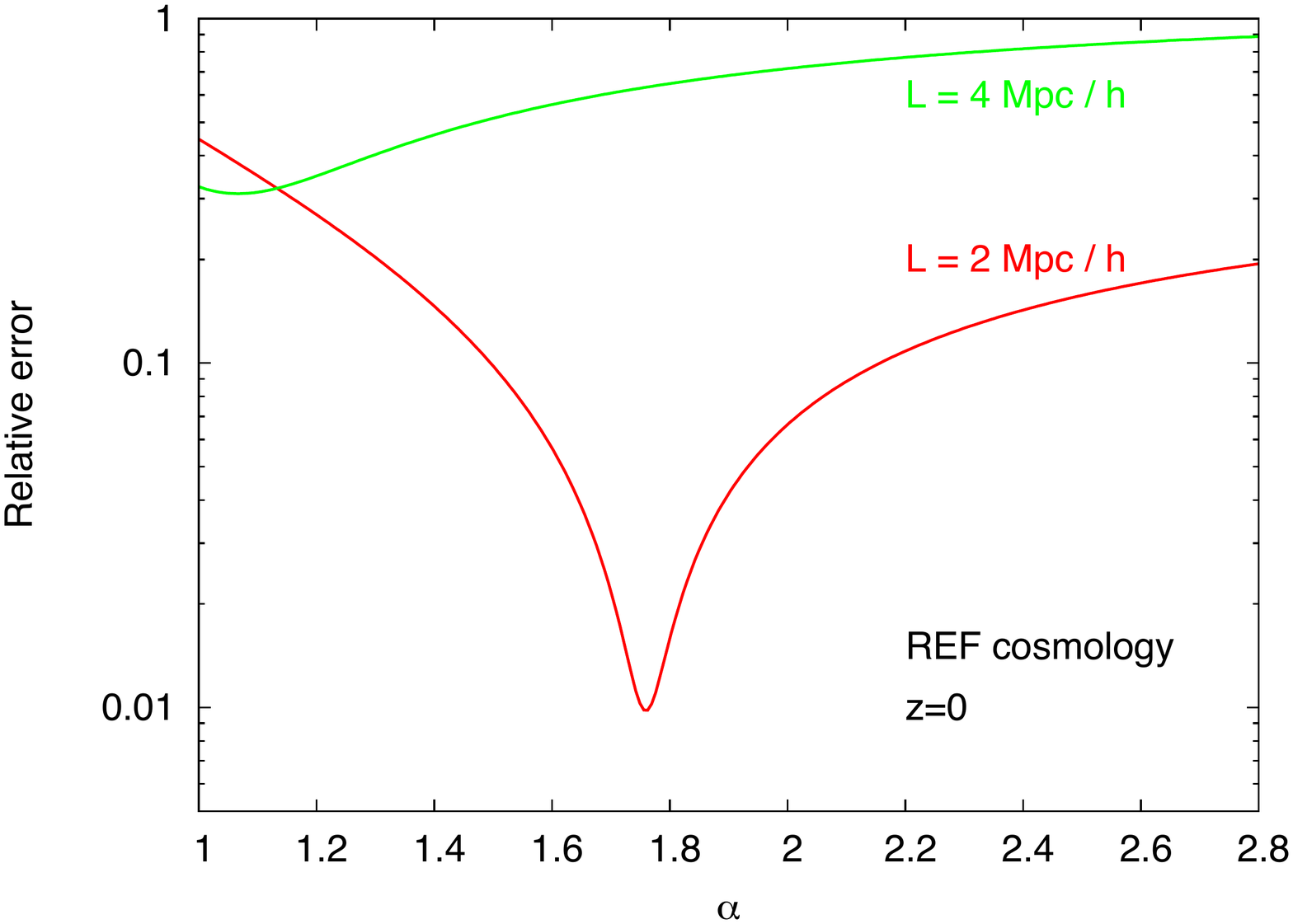}
\includegraphics[width=0.5\textwidth,angle=0]{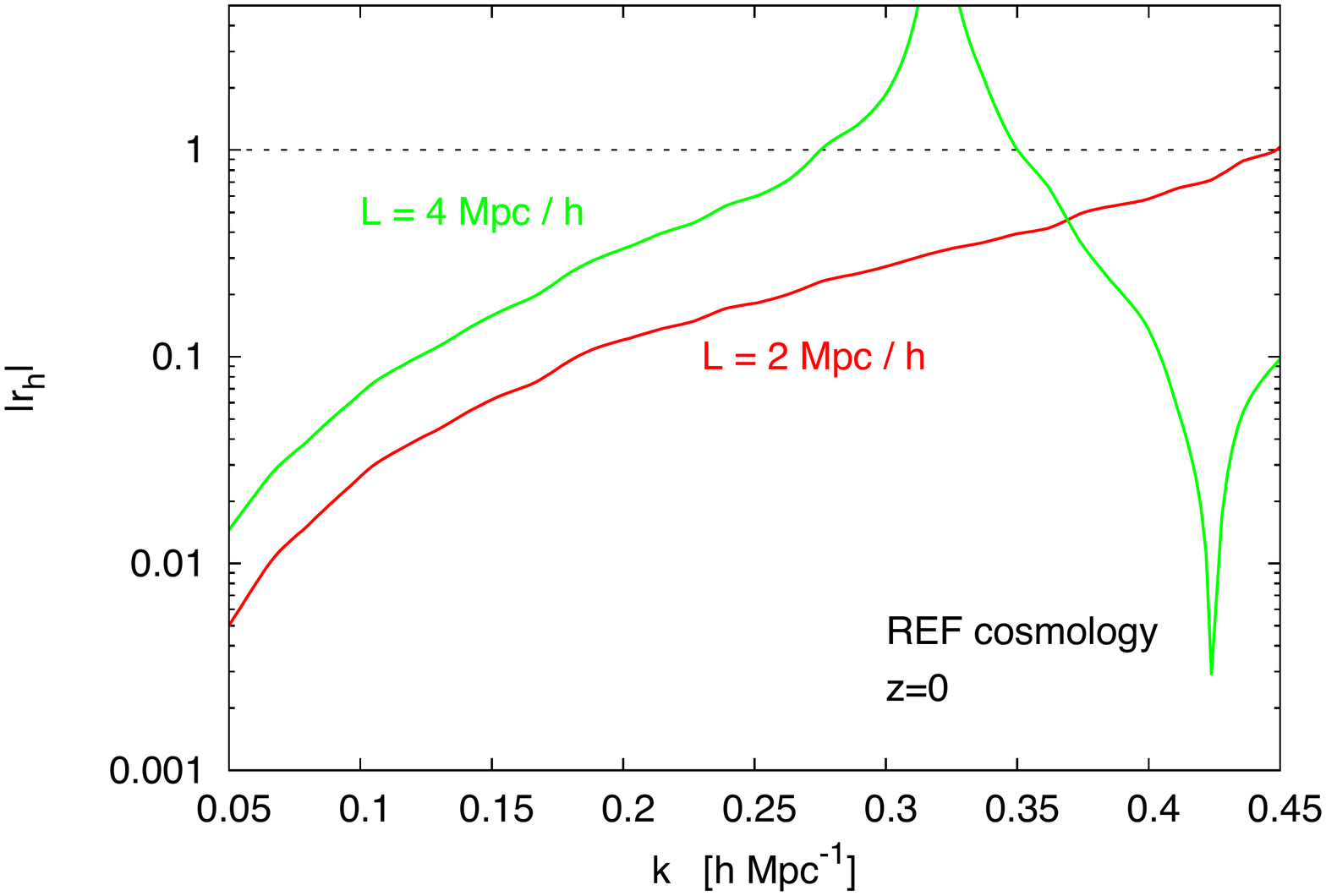}
}
\caption{Left panel: relative error  ${\cal E}$ in the PS reconstruction, given by eq. \re{error-E}, for two choices of $L$ in the filter function \ref{filter-app}. Right panel: Ratio \ref{rh} that provides an estimate of the deviation from SSA. See the text for details. 
}
\label{fig:compaL}
\end{figure*}

This is illustrated  the left panel of Figure \ref{fig:compaL}, which shows the relative error  ${\cal E}$ in the PS reconstruction procedure for the REF cosmology, $z=0$ and for the two choices   $L= 2  \;\mpch$ and  $L= 4  \;\mpch$. The relative error is computed according to eq.  \re{error-E}, as done in the main text.  We see from the figure  that the reconstruction  is significantly better in the   $L= 2  \;\mpch$ case. 

The reason for the poorer performance of the $L=4 \;\mpch$ filter can be understood by going back to our reconstruction formula, eq.~\re{P-reconstructed}. In practice, our procedure replaces  the SPT contribution to the correlator
\beq
\left\langle  h_2  \left( \bk, s \right)    \vpb_{1}(\bk',\eta) \right\rangle' \,, 
\eeq 
originating from loop momenta $q > k_{smooth}$ with the one measured from the simulation, which includes all nonlinear contributions. On the other hand, as discussed explicitly in the previous Appendix, the reconstruction formula computes the  
\beq
\left\langle  h_2  \left( \bk, s \right)    h_2(\bk',s') \right\rangle' 
\eeq
correlator in 1-loop SPT, with no other nonlinear contribution taken into account. In the right panel of Figure \ref{fig:compaL}  we plot the ratio
\begin{equation}
\!\!\!\!\!\!\!\! \!\!\!\!\!\!\!\! \!\!\!\!\!\!\!\! \!\!\!\!\!\!\!\! 
r_h \equiv   \frac{ 
\int_{\eta_{\rm in}}^\eta d s\, g_{12} \left( \eta - s \right) \int_{\eta_{\rm in}}^\eta d s' \,g_{12} \left( \eta - s' \right) \left( \langle h_2 \left( \bk, s \right)  h_2 \left( \bk, s' \right) \rangle'-\langle   h_{2} \left( \bk, s \right)  h_{2} \left( \bk, s' \right) \rangle'_{1-loop} \right)
}{
-  2\int_{\eta_{\rm in}}^\eta d s \,g_{12} \left( \eta - s \right)\left( \langle h_2 \left( \bk, s \right) \vpb_{1} \left( \bk, \eta \right) \rangle'-  \langle h_2 \left( \bk, s \right) \vpb_{1} \left( \bk, \eta \right) \rangle'_{1-loop}\right)
} \;,
\label{rh}
\end{equation} 
in which, both at the numerator and at the denominator we subtract the 1-loop counterpart from the correlator measured from the simulation. This ratio gives a measure of the importance of 2-loop and higher order contributions not included in the reconstruction formula, compared to those taken into account (we note that the denominator corresponds to the term that we have included in our reconstruction, and that gives rise to the second and third term at the right hand side of eq.  \re{rec2}). As we see, since the numerator scales as $k^4$ while the denominator as $k^2 P(k)$, the ratio grows at large $k$'s, which is the reason why the reconstruction formula fails for wave numbers larger than some limiting value, roughly related to the ratio becoming greater than unity. Moreover, the ratio increases with $L$, and therefore this limiting value is smaller for larger $L$. From the plot, it is clear that taking $L=4\;\mpch$  the 1-loop reconstruction formula fails already in the BAO region, while  $L=2\;\mpch$ preserves accuracy up to $k \simeq 0.4 \, \hmpc$, as we already saw in the main text. A next order computation,  will likely alleviate the filter dependence.

 \section*{References}
\bibliographystyle{JHEP}
\bibliography{/Users/pietroni/Bibliografia/mybib.bib}
\end{document}